\documentclass{article}

\usepackage[preprint]{neurips_data_2024}

\usepackage[utf8]{inputenc} 
\usepackage[T1]{fontenc}    
\usepackage{hyperref}       
\usepackage{url}            
\usepackage{booktabs}       
\usepackage{amsfonts}       
\usepackage{nicefrac}       
\usepackage{microtype}      
\usepackage{xcolor}         

\usepackage{graphicx}
\usepackage{subfigure}
\usepackage{wrapfig}
\usepackage{booktabs} 
\usepackage{array} 
\usepackage{multirow} 
\newcolumntype{L}[1]{>{\raggedright\arraybackslash}p{#1}}
\newcolumntype{R}[1]{>{\raggedleft\arraybackslash}p{#1}}

\usepackage{amsmath}
\usepackage{amssymb}
\usepackage{mathtools}
\usepackage{amsthm}

\usepackage{url}

\usepackage{epsfig} 
\usepackage{times} 
\usepackage{algorithm}
\usepackage{algpseudocode}

\usepackage{enumitem}
\setlist{nosep}

\usepackage{pifont}

\usepackage{float}
\usepackage{xspace}

\usepackage[capitalize,noabbrev]{cleveref}

\theoremstyle{plain}
\newtheorem{theorem}{Theorem}[section]
\newtheorem*{theorem*}{Theorem}

\newtheorem{example}[theorem]{Example}
\newtheorem{lemma}[theorem]{Lemma}

\theoremstyle{definition}
\newtheorem{definition}[theorem]{Definition}

\theoremstyle{remark}
\newtheorem{remark}[theorem]{Remark}

\usepackage[textsize=tiny]{todonotes}

\ifodd 1

\else

\newcommand{\ty}[1]{}
\newcommand{\zc}[1]{}
\newcommand{\kj}[1]{}

\newcommand{\com}[1]{}
\newcommand{\res}[1]{}

\fi

\newcommand{\PP}{\mathbb{P}}
\newcommand{\RR}{\mathbb{R}}
\newcommand{\wt}[1]{\widetilde{#1}}
\newcommand{\wh}[1]{\widehat{#1}}
\newcommand{\Prob}[1]{\PP\left(#1\right)}
\newcommand{\abs}[1]{\left\lvert#1\right\rvert}
\newcommand{\norm}[1]{\left\lVert#1\right\rVert}
\newcommand{\mc}[1]{\mathcal{#1}}
\newcommand{\ud}{\mathrm{d}\,}

\title{Measuring Fairness in Large-Scale Recommendation Systems with Missing Labels}

\author{%
    Yulong Dong \thanks{Both authors contributed equally to this research.}\\
    \texttt{david.dong@bytedance.com}\\
    TikTok Inc\\
    San Jose, CA, USA
    \And
    Kun Jin \thanks{Both authors contributed equally to this research.}\\
    \texttt{kun.jin1@bytedance.com}\\
    TikTok Inc\\
    San Jose, CA, USA
    \And 
    Xinghai Hu\\
    \texttt{xinghai.hu@bytedance.com}\\
    TikTok Inc\\
    San Jose, CA, USA
    \And
    Yang Liu\\
    \texttt{yang.liu01@bytedance.com}\\
    Bytedance Research\\
    San Jose, CA, USA
}

\begin{document}

\maketitle

\begin{abstract}
  In large-scale recommendation systems, the vast array of items makes it infeasible to obtain accurate user preferences for each product, resulting in a common issue of missing labels. Typically, only items previously recommended to users have associated ground truth data. Although there is extensive research on fairness concerning fully observed user-item interactions, the challenge of fairness in scenarios with missing labels remains underexplored. Previous methods often treat these samples missing labels as negative, which can significantly deviate from the ground truth fairness metrics. Our study addresses this gap by proposing a novel method employing a small randomized traffic to estimate fairness metrics accurately. We present theoretical bounds for the estimation error of our fairness metric and support our findings with empirical evidence on real data. Our numerical experiments on synthetic and TikTok's real-world data validate our theory and show the efficiency and effectiveness of our novel methods. To the best of our knowledge, we are the first to emphasize the necessity of random traffic in dataset collection for recommendation fairness, the first to publish a fairness-related dataset from TikTok and to provide reliable estimates of fairness metrics in the context of large-scale recommendation systems with missing labels.
\end{abstract}

\section{Introduction} \label{sec:intro}

Modern recommendation systems have achieved significant commercial success in a plethora of real-world domains. These systems aim to help users find items (e.g., videos, music, food, etc.) that are most relevant to their interests or search queries. However, there are rising concerns that these systems may have fairness issues, i.e., introduce biases against different stakeholders, e.g., certain groups of users or item creators \cite{Fair_Rec_Overview}. For example, non-mainstream music videos may be clicked at a lower rate than pop music videos, and thus recommendation models trained on the skewed data will likely recommend pop music videos more frequently to users that have similar interests to both categories, such imbalance can be further amplified by the recommendations without intervention, causing fairness issues \cite{Fair_Rec_Overview}. 

Our work focuses on the  \emph{Ranking-based Equal Opportunity (REO)} \cite{REO_paper} fairness notion, which measures utility-based, creator-side/item-side group fairness. Popular content creation systems, e.g., YouTube, Twitter, TikTok, and Kuaishou, specialize in personalized recommendations, rendering the \emph{preference label-dependent} REO fairness notion a more appropriate focus than the \emph{preference label-independent} fairness notions like \emph{Ranking-based Statistical Parity (RSP)} \cite{REO_paper}. The REO is an extension of the Equal Opportunity metric in binary classification \cite{Hardt_2016_NeurIPS} to the ranking/recommendation scenario, where REO requires the distribution of the recommendation outcome to be independent of the sensitive attribute when conditioned on the ground truth that the user is interested in the item. We will elaborate the metrics' definitions in \cref{sec:formulation}.

Designing trustworthy pipelines to compute the fairness metrics is vital for identifying fairness issues and building healthy and sustainable recommendation systems. However, most works have assumed that the true preference of a user in an item is known in the measurement, which is not the case in real-world situations. Modern large-scale recommendation systems face a persistent issue of missing labels for most of the user-item pairs. Specifically, for items that are never recommended to a user, the system will never observe the ground truth label. Moreover, these labels' information is necessary for accurate fairness metric measurements, and simplified approaches like treating missing labels as negative or choosing user item subsets with full labels can result in significantly misleading measurements. This is because the recommended items inherit the bias in the data, and the recommendation system learns this bias pattern. We use an example in \cref{sec:computation} to show the surprisingly misleading results when using these simplified approaches.

To address the issue of missing labels, our work proposes to utilize the random traffic data, which was originally introduced to detect the exposure bias \cite{Chen_2023_TIS}. We show that the random traffic data is an effective tool for us to accurately estimate user interest and successfully correct biases caused by simplified approaches. Moreover, in addition to globally accurate estimation of fairness metrics in large-scale recommendation systems, our work also studies the treatment effect in A/B tests through estimating and testing difference of fairness metrics (\cref{sec:fairness-monitoring}), which is important for monitoring fairness changes in new strategies. Towards this end, our work also derives a novel algorithmic paradigm for efficient statistical significance tests of fairness metrics with theoretical guarantees on the estimation errors. Compared to the permutation tests \cite{LinkedIn_Perm_Test} in literature, our novel methods have significantly improved computational advantage which paves the way for deployment and production on large-scale real-world commercial platforms. 

Parallel to the theoretical development of a novel framework for fairness estimation and benchmarking, we publish a real-world dataset from TikTok's short-form video recommendations, which, to our knowledge, is the first published dataset from TikTok for fairness study. As one of the largest short-video platforms, TikTok has over 1 billion monthly active users. In the dataset, we collected data from recommendation logs from Japan between April 18th 2024 and May 1st 2024. The data collection is performed by dumping logs uniformly at random. The default traffic dataset is collected on a daily basis where the daily data contains 150,000 rows of user-item pairs sampled uniformly from the recommendation logs using the default recommendation strategy. Given the relatively small size of random traffic, we sample 150,000 rows of user-item pairs uniformly at random from seven-day recommendation logs using a random recommendation strategy. We release engagement indices of these recommendation records and also an attribute derived from creators' age. In our published dataset, there are seven columns and $150,000 \times 14  = 2,100,000$ rows of default traffic data over 14 days and $300,000$ rows of random traffic data. We validate our framework by estimating and benchmarking fairness in our published dataset. Our experiments show the necessity of using random traffic, and the computational advantage and the correctness of our proposed methods.

At a high level, our work is closely related to \textbf{fairness in recommendation systems} \cite{Fair_Rec_Overview,User_Side_Fairness,REO_paper,Fair_Exp_Rank} as well as \textbf{random traffic in recommendation systems} \cite{KuaiRec,Chen_2023_TIS,Yang_2018_RecSys,Saito_2020_WSDM,Marlin_2009,Li_2010_WWW,gao2022kuairand}, please find more detailed discussions on related works in \cref{appsec:related}.

The main contributions of our work are three-fold. First, we identify the problem of existing literature when measuring fairness metrics in recommendation systems. When preference labels are missing in the large-scale recommendation dataset, we prove the necessity of including random traffic data for accurate estimations of label-dependent fairness metrics, and we show simplified approaches in previous literature fail. Second, we provide efficient and practical estimation algorithms as novel fairness benchmarking tools that monitor the fairness metrics, with theoretical guarantees on the error bounds. The theoretical analyses also provide new insights into the volume of traffic for accurate estimation. Finally, our numerical results on synthetic and TikTok's real-world data validate our theory and show the efficiency and effectiveness of our proposed method. Moreover, our work provides a carefully designed real-world dataset and a well-defined data collection procedure for creating recommendation datasets for fairness measurement.

\section{Ranking-based Equal Opportunity for Recommendation System} \label{sec:formulation}

Suppose our records in the recommendation system are user item pairs with $M$ user requests $\mathcal{U} := \{u_1, u_2, \dots, u_M\}$ (some user requests may correspond to the same physical user), and $N$ items $\mathcal{I} := \{i_1, i_2, \dots, i_N\}$. Our data set $\mathcal{D}$ consists of $M \times N$ rows (we show the ideal case for clarity of explanation and derivation first, and we will elaborate on the realistic setting later), each row corresponding to a user-item pair $(u_m, i_n)$. Specifically, each row looks like
\begin{equation*}
    \left(~ u_m, i_n, R(u_m, i_n), Y(u_m, i_n), S(u_m, i_n) ~\right),
\end{equation*}
where $R(u_m, i_n) \in \{0,1\}$ indicates the actual \textbf{recommendation decision} by the recommendation system (the default traffic, which we will elaborate on later), $R(u_m, i_n)=1$ (resp. $R(u_m, i_n)=0$) indicates $i_n$ is recommended (resp. not recommended) to the request $u_m$; $Y(u_m, i_n) \in \{0,1\}$ indicates the actual \textbf{preference/relevance label}, $Y(u_m, i_n)=1$ (resp. $Y(u_m, i_n)=0$) indicates $i_n$ is relevant (resp. not relevant) to the request $u_m$; $S(u_m, i_n) \in \mathcal{S}$ denotes the \textbf{sensitive attribute} of the pair $(u_m, i_n)$ and $\mathcal{S} = \{s_1, \dots, s_K\}$ is the set of sensitive attributes ($K := |\mathcal{S}|$). These sensitive attributes partition the entire set of user-item pairs into disjoint groups, and we call group $k$ as the group of pairs with sensitive attribute $s_k$. 

In an early work \cite{REO_paper}, the REO was introduced as an item-side/creator-side fairness notion, namely $S(u_m, i_n) = S(i_n)$, where the user-dependency is dropped. We note that our results and analyses are applicable to the general setup by viewing them as random variables defined on the sample space consisting of all user-item pairs, which does not necessitate any dropping of dependency. For notational simplicity, we hide the $(u, i)$ dependency in these random variables and simplify the tuple as $(u, i, R, Y, S)$ without confusion. 

The REO fairness measures the disparity of ranking-based true positive utilities between the groups. Formally, the Ranking-based true positive rate (RTPR) utility of group $k$ is defined as 
\begin{equation} \label{eq:REO_utility}
    U_k  := \mathbb{P}(R=1 | Y=1, S=s_k).
\end{equation}
Under the settings in \cite{REO_paper}, $U_k$ stands for the probability that a user gets a recommended item created by creators from the $k$-th group when the user has a positive preference for it. At a high level, REO is a derivative of the Equal Opportunity (EO) fairness notion that fits the ranking setting, where $R=1$ is considered a ``positive prediction''.
The REO fairness penalty is 
\begin{equation} \label{eqn:REO_penalty}
    \Delta_\mathrm{REO} := \frac{\mathrm{std}(U_1, \dots, U_K)}{\mathrm{mean}(U_1, \dots, U_K)}.
\end{equation}

While the fairness penalty $\Delta_\mathrm{REO}$ is a reasonable global fairness measurement, it can not directly tell which group is advantaged or disadvantaged. Therefore, for a given group, we define the relative group utility as 
\begin{equation} \label{eqn:delta_Uk}
    \Delta U_k := \frac{U_k}{\mathrm{mean}(U_1,\dots,U_K)} - 1,
\end{equation}
which measures the deviation of the group's utility to the mean value. Its sign stands for the advantage or the disadvantage of the group. 
Throughout the paper, these quantities, including group utilities, relative group utilities and fairness penalty, are referred to as \textit{REO related metrics} without otherwise noted when the context is clear.

\section{Computation of REO Related Metrics} \label{sec:computation}

In this part, we present realistic barriers in computing the REO related metrics and propose using random traffic to come up with consistent estimations.

\subsection{Identifiability of REO}\label{sec:identifiability-overview}
The full set of user-item pairs can be partitioned into two subsets, one containing pairs that are recommended ($R=1$, recommended subset) and the other with pairs that are not ($R=0$, unrecommended subset). In the recommended subset, the users' engagement actions (e.g., like, share, etc.) can reflect their preference labels $Y$ on the items. In contrast, those labels on the unrecommended subset remain unknown. Due to the uncertainty in the unrecommended subset, REO metrics are actually \textbf{not identifiable} from the partially observed subset \textbf{without probing the unrecommended subset}. To illustrate, we provide a simple numerical example as follows.

\begin{example} \label{example:not_identifiable}
    Consider two datasets derived in a recommendation system that contains two sensitive attributes $s_1$ and $s_2$. Aggregating the user-item pair count according to the recommendation decision and preference label, these datasets become the following pivot table.
\begin{table}[ht]
\centering
\begin{tabular}{c|cc|cc}
        \hline
        & \multicolumn{2}{c|}{dataset A} & \multicolumn{2}{c}{dataset B} \\ \hline
        & $Y = 0$       & $Y = 1$        & $Y = 0$      & $Y = 1$        \\ \hline
$R = 0$ & 100,000 / 100,000       & 0 / 0          & 99,900 / 100,000        & \textcolor{red}{100} / 0        \\
$R = 1$ & 0 / 0         & \textcolor{orange}{100} / \textcolor{teal}{100}      & 0 / 0        & \textcolor{blue}{100} / \textcolor{purple}{100}     \\ \hline
\end{tabular}
\vspace{1em}
\caption{A toy example demonstrating the unidentifiability of REO metric due to missing labels. The first number before the slash in each entry stands for the value of attribute $s_1$ and that after the slash stands for the value of attribute $s_2$, (e.g., 100/0 means 100 pairs in group 1 and 0 pair in group 2).}
\end{table}\\
On dataset A, a direct computation gives $U_1 = \Prob{R = 1 | Y = 1, S = s_1} = \textcolor{orange}{100}/\textcolor{orange}{100} = 1$ and $U_2 = \Prob{R = 1 | Y = 1, S = s_2} = \textcolor{teal}{100}/\textcolor{teal}{100} = 1$. Hence, the dataset is perfectly fair as $\Delta_\mathrm{REO} = 0$. However, on dataset B, it can be computed that $U_1 = \textcolor{blue}{100} /(\textcolor{blue}{100}+\textcolor{red}{100}) = \frac{1}{2}$ and $U_2 = \textcolor{purple}{100}/\textcolor{purple}{100} = 1$. Consequently, $\Delta_\mathrm{REO} = 2 / 3$. It is worth noting that these two datasets completely agree on the recommended subset where $R = 1$. Hence, without any information from the unrecommended subset where $R = 0$, they are not distinguishable but belong to completely different fairness. 
\end{example}

\begin{lemma}[informal]\label{lem:exist-fair-and-unfair-informal}
    In terms of REO, there always exists a perfectly fair ($\Delta_\mathrm{REO} = 0$) set of user-item pairs dataset and an unfair ($\Delta_\mathrm{REO} > 0$) set of user-item pairs such that they agree on the recommended subset.
\end{lemma}

A formal version of the proceeding lemma and its proof is given in \cref{app:identifiability}. Based on the existence of these counterparts, we are able to prove the unidentifiability of REO metrics as in the following theorem. We note that our definition of REO metrics being {not identifiabile} is that there always exist two datasets that fully agree on the recommended subset but their REO metrics are not able to be computed accurately at the same time by any algorithm (a comprehensive discussion is presented in \cref{app:identifiability}). Conversely, a quantity is identifiable means that there exists an algorithm so that for any input dataset of a fixed size, it estimates that quantity with uniformly small estimation error with high probability. 

\begin{theorem}[informal]\label{thm:identifiability}
    The REO metrics are generally not identifiable without probing the unrecommended subset.
\end{theorem}

In \cref{app:identifiability}, we prove a formal version of the identifiability in \cref{thm:identifiability-rigorous}. The above theorem reveals the importance of probing and storing information inside the unrecommended subset in order to compute REO. The probing strategy we proposed in this work is referred to as \emph{random traffic}, which is a \emph{uniform sampling} procedure on the user-item pairs. 

This part shows the necessity of random traffic data (probes to the unrecommended subset) for accurate estimation of REO metrics. But a natural question is \textbf{``Can we use a model to predict user preferences to get accurate estimations and bypass the random traffic data?''}. Unfortunately, the answer is \textbf{NO}. While using the actual model performance on the entire dataset can accurately estimate the REO related metrics \cite{Zhu_2023_ICML, Amazon_2022_FAccT}, using the estimated model performance from a biased subset can not. Intuitively, without the random traffic, we can not accurately evaluate the model's performance on the entire dataset. Similar to \cref{example:not_identifiable}, given a dataset, there exist two models that have identical predictions on the recommended subset, and thus the same estimated model performance. 
However, these two models can have significantly different prediction performances on the unrecommended subset since its distribution is significantly different from the recommended subset. In this case, the two models can have significantly different actual model performances on the entire dataset. As shown in \cref{app:identifiability}, the use of random traffic is inevitable to overcome the uncertainty due to the counterfactual nature, which is also applicable to the use of arbitrary machine learning models.
As the utility function is factored into components measurable from random and default traffic through explicit formulas, training a perfect model using these traffics is equivalent to directly estimating metrics. Moreover, as the data changes over time in recommendation systems, we need recent data to re-train or re-evaluate the model's performance, and thus the need for random traffic persists.

\subsection{Random Traffic Data}
\begin{figure*}[h!]
    \centering
    \begin{minipage}[b]{.5\textwidth}
        \centering
        \includegraphics[width=.8\linewidth]{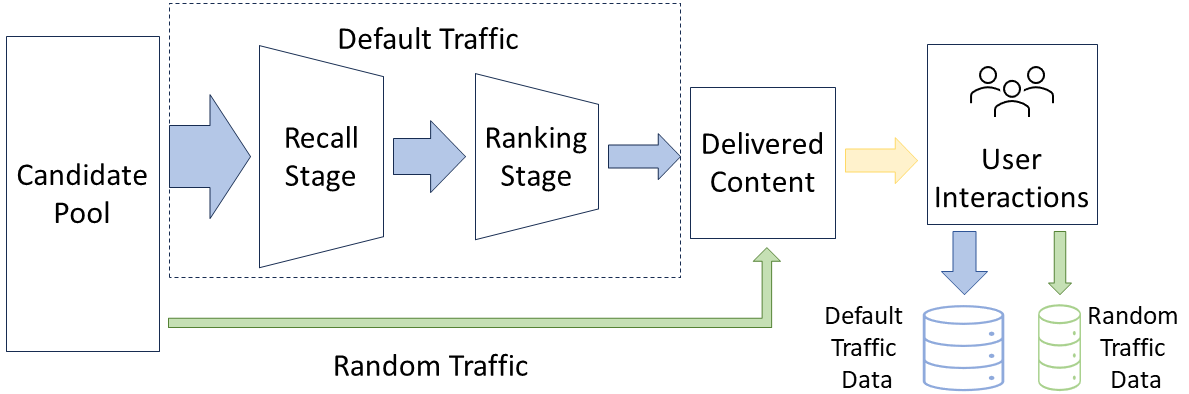}
        \caption{An illustration of Random Traffic through forced insertion and Default Traffic through the default recommendation strategy, which generates $\mathcal{D}_{rand}$ and $\mathcal{D}_{rec}$ respectively.}
        \label{fig:random_traffic}
    \end{minipage}
    \hfill
    \begin{minipage}[b]{.47\textwidth}
        \centering
        \includegraphics[width=.8\linewidth]{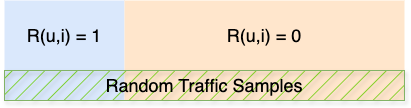}
        \caption{An illustration of uniform sampling  in Random Traffic $R(u,i)=1$ (resp. $0$) means $i$ is recommended (resp. not recommended) in the default traffic for a user request $u$.}
        \label{fig:random_traffic_samples}
    \end{minipage}
\end{figure*}

To estimate components causing the counterfactual nature of REO metrics as discussed in the previous subsection, we introduce the \emph{random traffic}, which is independent of the \emph{default traffic} throughout this paper. \cref{fig:random_traffic} illustrates the idea of default and random traffic. To generate the random traffic, we first determine whether we activate random sampling for every incoming request, where the activation follows a Bernoulli distribution with an activation probability of $p_{act} > 0$, which is usually a sufficiently small number (e.g., $p_{act} < 10^{-3}$). If random sampling is not activated, the request receives recommendations entirely from the default traffic. On the other hand, if random sampling is activated for a request, we uniformly at random choose items from the whole candidate pool to recommend (random traffic can recommend the same candidates as the default traffic, de-duplication is discussed in \cref{appsec:dedup}), and \cref{fig:random_traffic_samples} shows the sampling logic. We denote $\mathcal{U}_{rec}, \mathcal{I}_{rec}, \mathcal{D}_{rec}$ (resp. $\mathcal{U}_{rand}, \mathcal{I}_{rand}, \mathcal{D}_{rand}$) as the corresponding set of user requests, items, and user-item pairs in the default traffic (resp. random traffic). 

In actual production, companies may only maintain log data for user-item pairs that are indeed recommended to the users. Since the full daily log data contains an extremely large amount of user-item information (e.g., $> 10^8$ rows per day), giant content creation platforms may sample uniformly at random from such user-item pairs and dump them to the log data to reduce the computation and memory cost. Throughout this paper, we assume that user-item pairs in $\mathcal{D}_{rec}$ are sampled uniformly at random from user-item pairs in $\mathcal{D}$ such that $R(u,i)=1$. We note that while $\mathcal{D}_{rec}$ is always stored, $\mathcal{D}_{rand}$ is only available when random traffic is implemented. We show the necessity of random traffic in \cref{thm:identifiability-rigorous} and below we provide analysis given that random traffic data is available.

\subsection{Fairness Metrics and Fairness Monitoring}

In this part, we discuss the actual application of REO fairness monitoring. We outline a high-level overview of our computation, and a more detailed discussion is referred to \cref{sec:fairness-monitoring}.

Note that the definition of the group utility in \cref{eq:REO_utility} involves counterfactual events as the users' preference label is identified after they receive the recommended item. The causality renders the direct evaluation of the group utility infeasible in accordance with the identifiability argument in the previous subsection. To compute REO related metrics, we need to factor the group utility into measurable components by leveraging default and random traffic. Using the Bayesian theorem, we have 
\begin{equation*}
    U_k = \Prob{R = 1 | Y = 1, S = s_k} = \frac{\Prob{Y = 1, S = s_k | R = 1}}{\Prob{Y = 1, S = s_k}} \Prob{R = 1}.
\end{equation*}
Then, the group utility is assembled from three probabilities, which are either measurable or irrelevant as discussed in the paragraph below. From the random and default traffic, we are able to estimate the following quantities:
\begin{equation*}
    \hat{P}_{k} := \frac{\sum_{(u, i) \in \mathcal{D}_{rand}} \mathbb{I}(Y(u, i)=1) \mathbb{I}(S(i) = s_k)}{|\mathcal{D}_{rand}|}, \hat{Q}_{k} := \frac{\sum_{(u, i) \in \mathcal{D}_{rec}} \mathbb{I}(Y(u, i)=1) \mathbb{I}(S(i) = s_k)}{|\mathcal{D}_{rec}|}.
\end{equation*}
Consequently, REO metrics are reassembled as follows:
\begin{equation} \label{eqn:REO_penalty_estimation}
    \hat{U}_k := \frac{\hat{Q}_k}{\hat{P}_k},\ \wh{\Delta U_k} := \frac{\hat{U}_k}{\mathrm{mean}(\hat{U}_1,\dots,\hat{U}_K)} - 1, \text{ and } \wh{\Delta_\mathrm{REO}} := \frac{\mathrm{std}(\hat{U}_1, \dots, \hat{U}_K)}{\mathrm{mean}(\hat{U}_1, \dots, \hat{U}_K)}.
\end{equation}
It is worth noting that the estimator $\hat{U}_k$ is equal to the utility function up to a scalar $\Prob{R = 1}$ which is independent of the group index ($k$). Note that the relative group utility and the fairness penalty are invariant under the simultaneous rescaling of utility functions. Hence, this difference in the estimator will not affect the estimation of the metric of interest. The rescaling scalar $\Prob{R = 1}$ stands for the probability that a random user-item pair goes through a recommendation process. Due to the extremely large size of the possible user-item pairs and the large volume of the recommendation log, exactly measuring this quantity is highly infeasible. However, its appearance in the utility function is inevitable due to the counterfactual nature. Thanks to the Bayesian theorem, we factor the utility function as two measurable quantities using default and random traffic, and we get rid of the unmeasurable probability according to the scale invariance.

Recall, \cref{sec:identifiability-overview} shows that REO metrics are not identifiable without probing the unrecommended subset using e.g. random traffic. In this subsection, with random traffic, we are able to prove that our proposed estimators consistently estimate REO metrics with sufficient traffic size. The performance is guaranteed by the following theorem. A series of more comprehensive statements and the proof are presented in \cref{sec:proofs}.

\begin{theorem}[informal]\label{thm:main_theorem_estimation}
    Suppose the traffic sizes are sufficiently large $\abs{\mathcal{D}_{rec}}, \abs{\mathcal{D}_{rand}} = O\left( K^2 \epsilon^{-2} \log\left(K \delta^{-1}\right)\right)$. Then with probability at least $1 - \delta$, the estimation errors are uniformly upper bounded:
    \begin{equation*}
        \max_{k = 1, \cdots, K} \abs{\wh{\Delta U_k} - \Delta U_k} \le \epsilon \text{ and } \abs{\wh{\Delta_\mathrm{REO}} - \Delta_\mathrm{REO}} \le \epsilon.
    \end{equation*}
\end{theorem}

Besides the guaranteed accuracy of estimation, we also developed an extremely simple method for quantifying the confidence intervals of the estimators of fairness metrics which is outlined in \cref{alg:delta-method} leveraging the statistical distribution of REO metrics and delta method \cite{Doob1935}. In A/B tests, the treatment effect is quantified as the difference in fairness metrics, e.g. $D_\mathrm{REO} = \Delta_\mathrm{REO}^\mathrm{treatment} - \Delta_\mathrm{REO}^\mathrm{control}$. We also provide fast and reliable significance tests for treatment effects in A/B tests in \cref{alg:delta-method-AB}. These pave the way for deploying our fairness monitoring and benchmarking framework in large-scale real-world commercial platforms. In the later section, we will showcase the numerical results in TikTok's recommendation dataset, which is published alongside this paper.

\section{Numerical Results} \label{sec:numerical}
\subsection{Synthetic Data and Numerical Tests}

To validate our fairness metrics computation, we numerically perform the computation on synthetic data. We assume there are two groups in the study and define the proportion of each group as follows. Specifically, we create 3 sets of data, where on the random traffic, the relevant proportions are (1) $p_1 = 0.01, p_2 = 0.05$; (2) $p_1 = 0.001, p_2 = 0.005$; (3) $p_1 = 0.0001, p_2 = 0.0005$, and we simulate other relevant proportions as follows.
On default traffic, the relevant proportions are set to $q_1 = 10 \times p_1$ and $q_2 = 5 \times p_2$, where we assume the recommendation algorithms are effective and thus $q_k > p_k$. The meaning of these probabilities are $p_k = \Prob{Y = 1, S = s_k}$ and $q_k = \Prob{Y = 1, S = s_k | R = 1}$. To assemble a valid normalized probability vector, we also need their complements. For simplicity, we set $\tilde{p}_1 := \Prob{Y = 0, S = 1} = 0.25 \times (1 - p_1 - p_2)$ and $\tilde{p}_2 := \Prob{Y = 0, S = 2} = 0.75 \times (1 - p_1 - p_2)$. The same procedure is performed to proportions $\wt{q}_1, \wt{q}_2$ on default traffic. These form two valid probability vectors standing for the scenario with infinitely many samples. 
Then, for a given traffic size $n$, the data generation process is emulated by drawing multinomial random variables $(p_1(n), p_2(n), *, *) \sim \mathrm{Mult}(n, (p_1, p_2, \tilde{p}_1, \tilde{p}_2))$ and $(q_1(n), q_2(n), *, *) \sim \mathrm{Mult}(n, (q_1, q_2, \tilde{q}_1, \tilde{q}_2))$. The fairness penalty $\Delta_\mathrm{REO}(n)$ is computed with these sampled versions which contain effects of finite sample size. The ground truth $\Delta_\mathrm{REO}$ is set to the fairness penalty computed from the predefined portions. For each $n$, we independently repeat the sampling and computation for 50 times to get the error bar. 
    \begin{wrapfigure}{l}{0.45\textwidth} 
        \includegraphics[width=.425\textwidth]{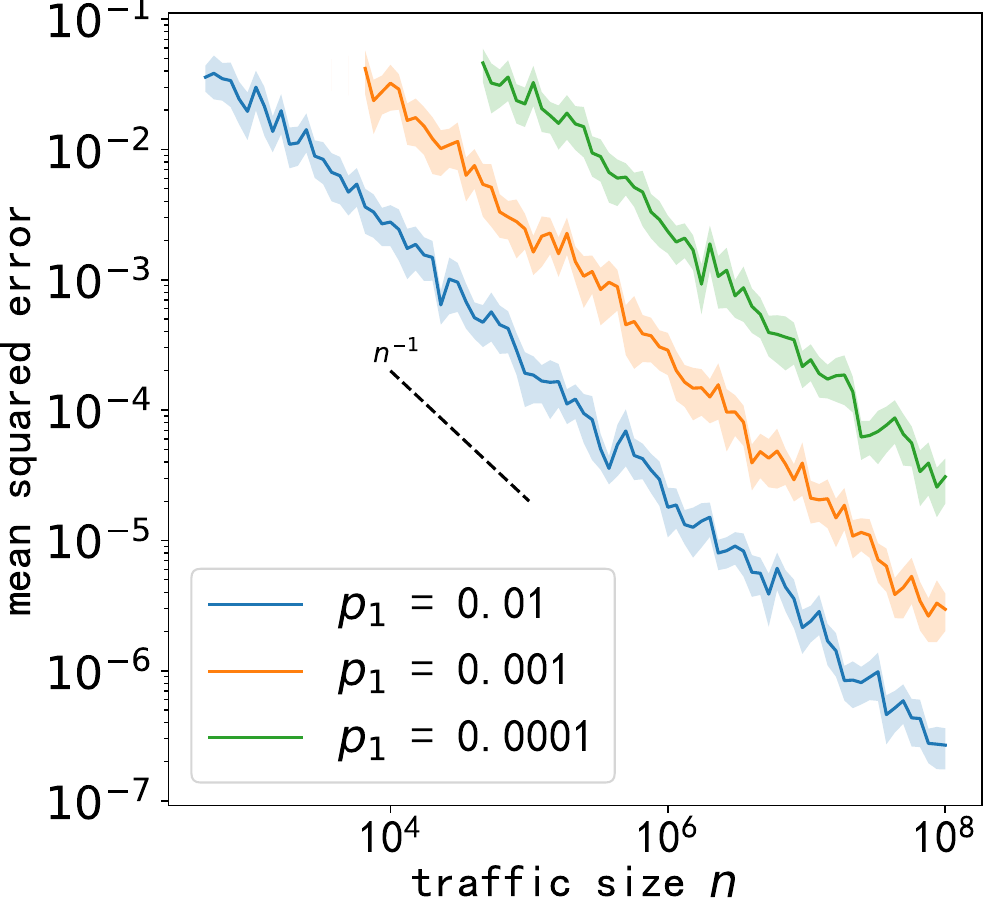}
        \caption{The mean squared error of fairness penalty on synthetical data with variable traffic size.}
        \label{fig:synthetical_test}
    \end{wrapfigure}
The mean squared error $\mathrm{MSE}(n) := \mathbb{E}((\Delta_\mathrm{REO}(n) - \Delta_\mathrm{REO})^2)$ as a function of sample size $n$ is displayed in \cref{fig:synthetical_test}. By comparing with the dashed line, we see the scaling of the curve is $\mathrm{MSE}(n) \sim 1 / n$. Note that this is essentially the variance scaling according to the law of large numbers. Hence, according to the bias-variance decomposition of MSE, the result suggests that the bias in the estimator is almost vanishing. Furthermore, we see that when the relevant portions gets smaller, the estimation problem becomes increasingly challenging where the MSE of the same sample size is much higher. This validates our theoretical proof.

\subsection{Real-World Data and Description}
In this part, we present our numerical results on the real-world data collected from the recommendation logs of the video-sharing platform TikTok. The data contains regular traffic obtained using the default recommendation strategy, which corresponds to our $\mathcal{D}_{rec}$, as well as the random traffic data $\mathcal{D}_{rand}$. We refer the detailed specification of the dataset information and collection method to \cref{sec:intro}. There are seven boolean fields in the dataset whose meanings are outlined as follows. The first six fields are used as positive signals to indicate the user's preference. When any one among these six signals is positive, we have $Y(u,i)=1$ for this user-item pair. The last field is referred to as the sensitive attribute of the video creator in the study.
\begin{enumerate}[leftmargin=*,label=(\arabic*)]
    \item \textbf{like\_video} stands for whether the user clicked the ``like'' button on this video.
    \item \textbf{share} means if the user shared the video internally or externally.
    \item \textbf{follow} stands for whether the user started to follow the creator after viewing the video. Note that it means that the user is not a follower of the creator at the time of viewing the video.
    \item \textbf{finish} means if the user finished viewing the full video.
    \item \textbf{download} stands for whether the user downloaded the video after viewing it.
    \item \textbf{long\_view} indicates whether the user stayed on watching the video for a sufficiently long time.
    \item \textbf{young\_adult} indicates whether the video creator's age is between 18 and 24 (both inclusively).
\end{enumerate}

\begin{figure*}[h!]
    \begin{minipage}[b]{.5\textwidth}
        \centering
        \includegraphics[width=.65\linewidth]{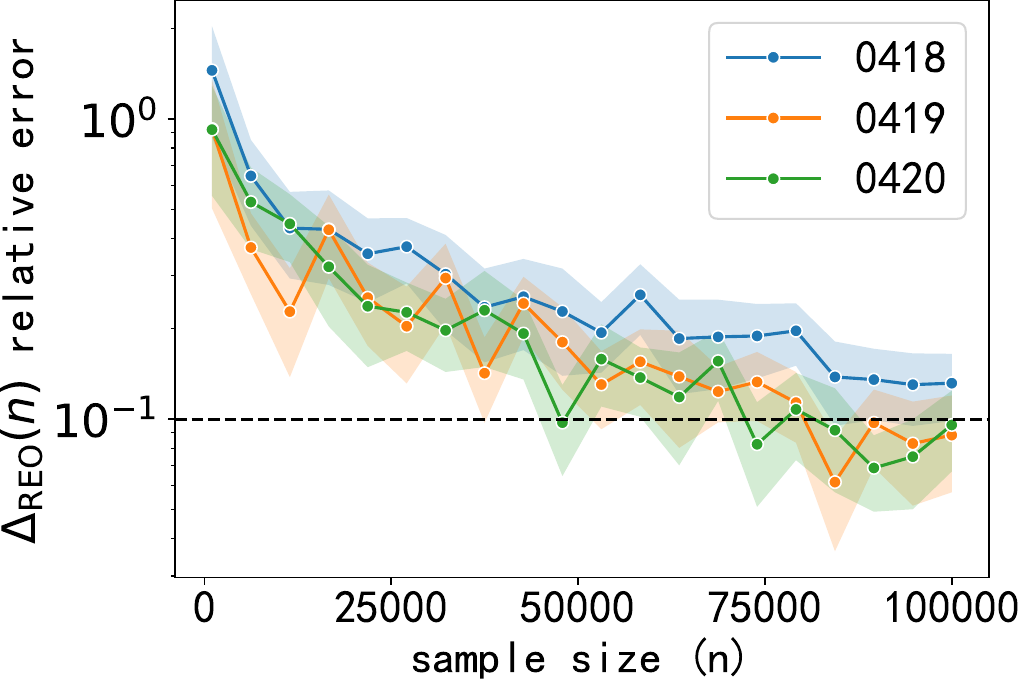}
        \caption{Relative error of fairness penalty estimation as a function of dataset size.}
        \label{fig:sample-size}
    \end{minipage}
    \hfill
    \begin{minipage}[b]{.47\textwidth}
        \centering
        \includegraphics[width=\linewidth]{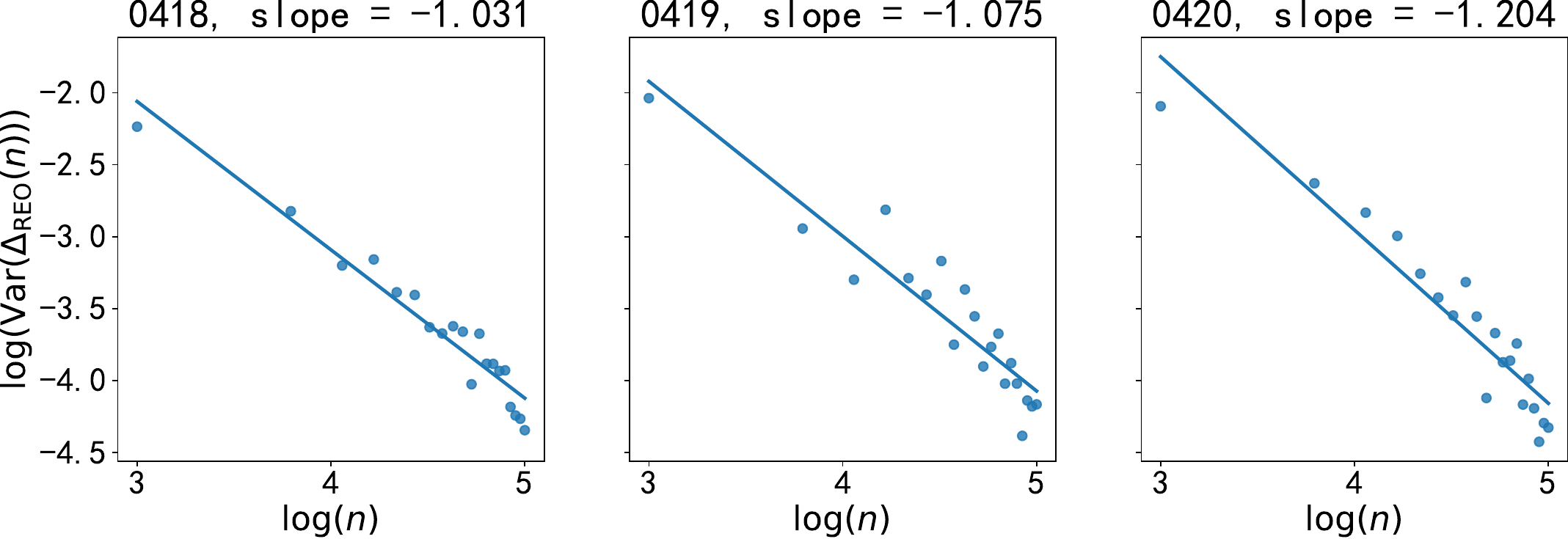}
        \caption{Variance of fairness penalty estimation as a function of dataset size.}
        \label{fig:sample-size-variance}
    \end{minipage}
\end{figure*}

We numerically illustrate the computation accuracy of the fairness penalty with variable sample sizes. We set the the metric value with a full 150,000 rows of daily data as the benchmark $\Delta_\mathrm{REO}$. We then sample uniformly from the full dataset with a given sample size $n$ and compute the metric value $\Delta_\mathrm{REO}(n)$ with this subset. The accuracy is quantified as the relative error $\abs{\Delta_\mathrm{REO}(n) - \Delta_\mathrm{REO}} / \Delta_\mathrm{REO}$. For each $n$, the procedure is independently repeated 20 times. \cref{fig:sample-size} shows a rapid decrease in the relative error when the sample size gets larger. In \cref{fig:sample-size-variance}, we also visualize the variance of $\Delta_\mathrm{REO}(n)$ as a function of sample size $n$. By performing linear regression after logarithmic transformations, we observe the scaling of variance is close to $1/n$ which agrees with our theoretical analysis.

\subsection{Within-strategy fairness monitoring}
    \begin{wrapfigure}{l}{0.5\textwidth} 
        \includegraphics[width=.45\textwidth]{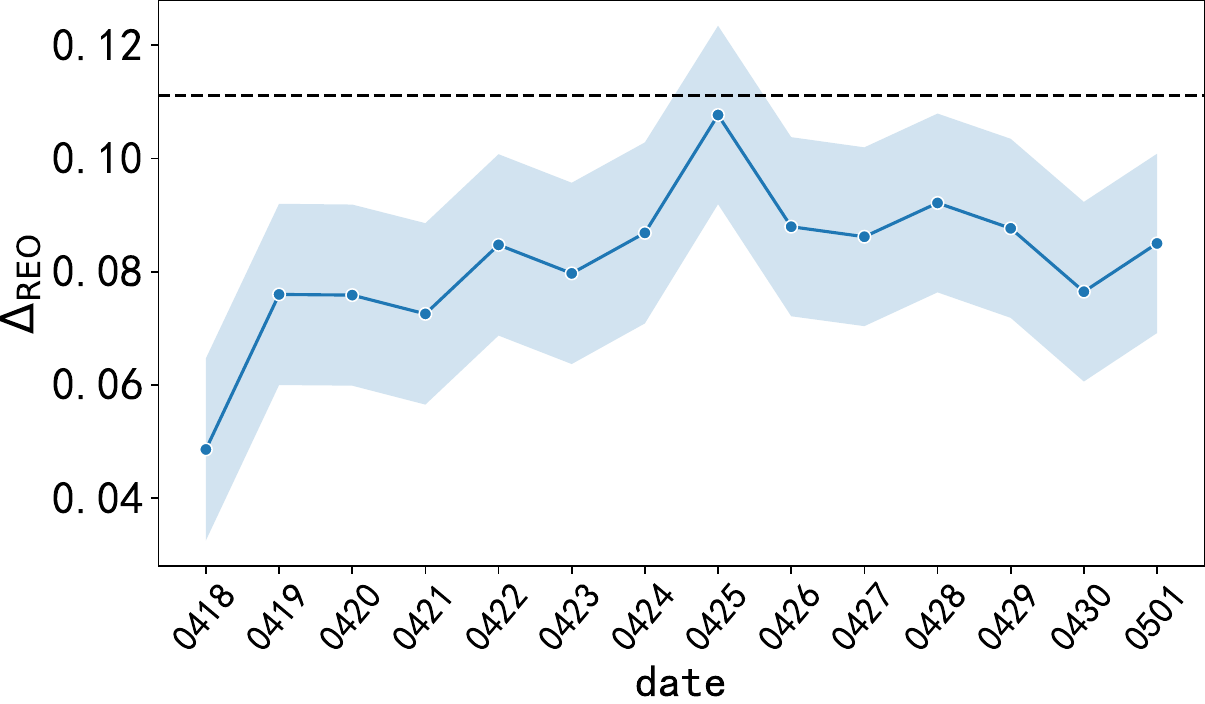}
        \caption{The fairness penalty for young\_adult attribute in a two-week window. The dashed line marks the threshold of 11.1\%.}
        \label{fig:Delta_REO_JP_0418_0501}
    \end{wrapfigure}

We monitor the daily fairness penalty in this dataset by deriving the estimated value and its confidence interval according to \cref{alg:delta-method}. The result is visualized in \cref{fig:Delta_REO_JP_0418_0501}. It reveals that except for the value on April 25th, the daily fairness penalties are significantly lower than the threshold of $11.1\%$, which is a requirement for the 80$\%$ rule to hold \cite{Feldman_2016_KDD} (when $U_1 = 0.8 U_2$, $\mathrm{std}(U_1, U_2)/\mathrm{mean}(U_1, U_2) = 1/9 = 11.1\%$). 
The increase in the fairness penalty might be caused by some tests of online strategies. However, the fairness penalty value is lower than the threshold and the results indicate that the breaking of the threshold is not significant. 

\begin{figure*}
    \begin{minipage}[b]{.45\textwidth}
    \centering
    \includegraphics[width=\linewidth]{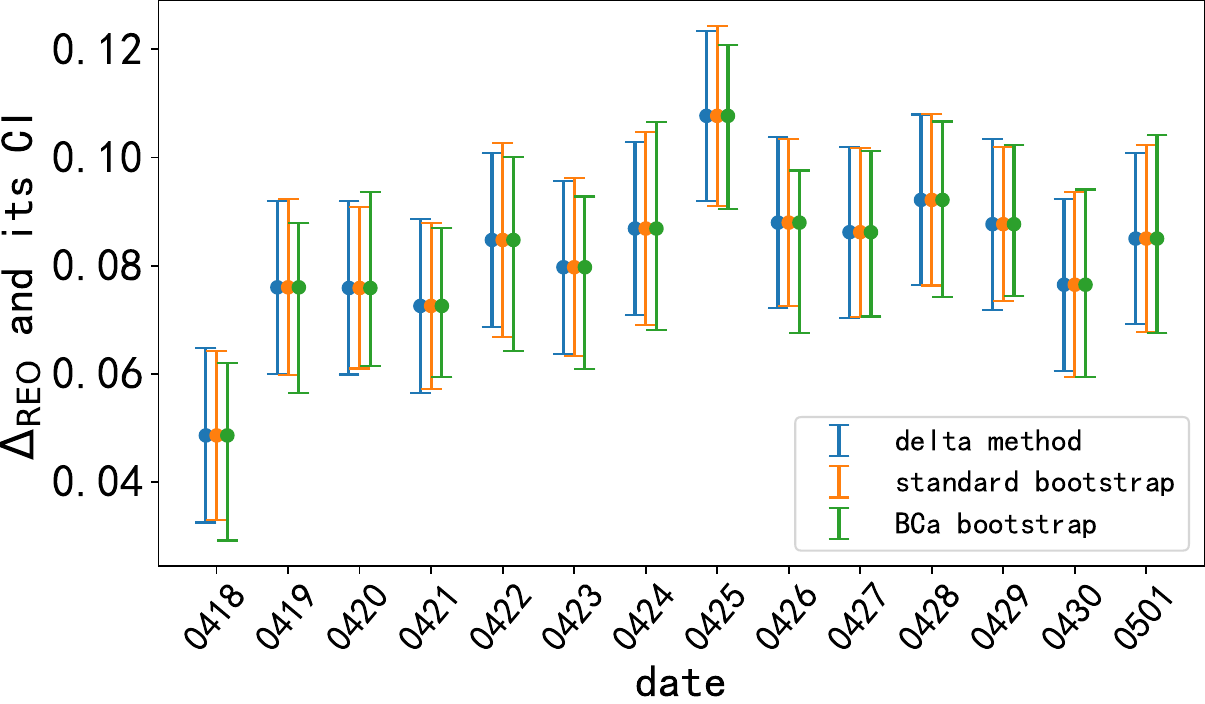}
        \caption{Comparing the confidence interval (CI) derived from multiple methods.}
        \label{fig:confidence_interval_JP_0418_0501}
    \end{minipage}
    \hfill
    \begin{minipage}[b]{.45\textwidth}
    \centering
    \includegraphics[width=\linewidth]{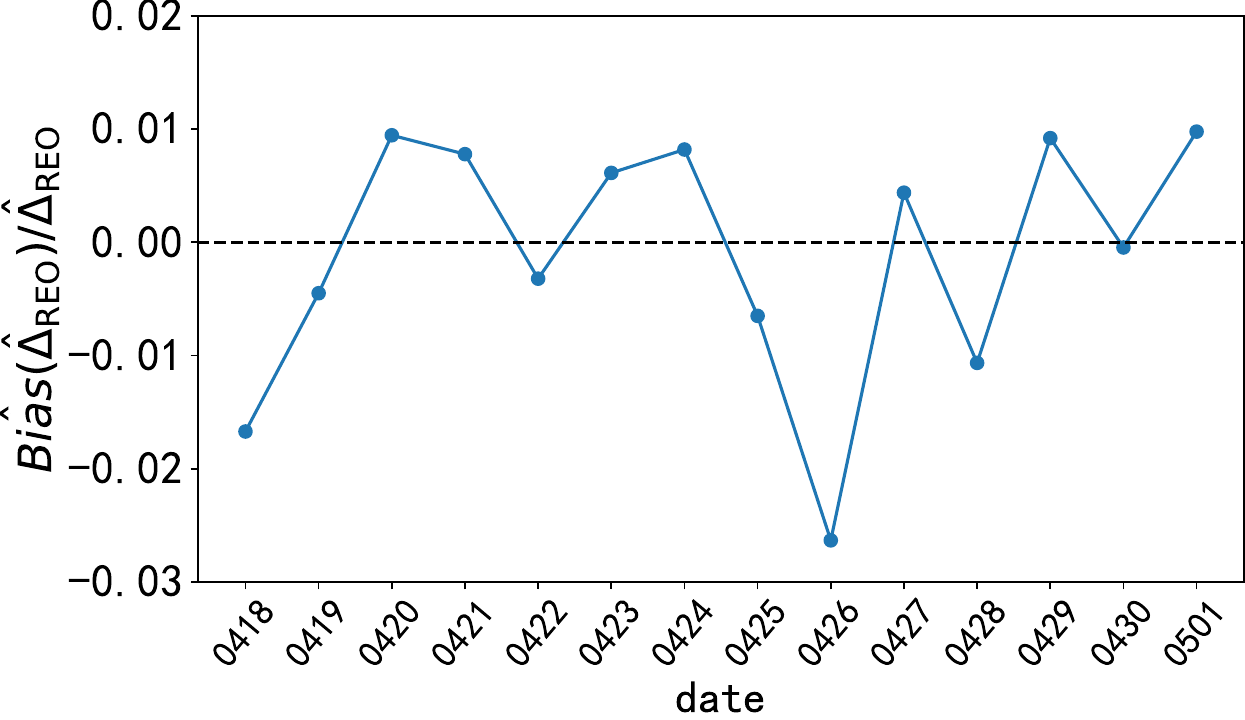}
        \caption{Bias of the estimator of REO metric derived from bootstrap method.}
        \label{fig:bootstrap-bias}
    \end{minipage}
\end{figure*}

We perform numerical tests to validate our new delta method based significance test in \cref{alg:delta-method}. We derive the confidence interval for each daily traffic by using three methods: delta method, standard bootstrap and bias-correction accelerated (BCa) bootstrap. Both bootstrap methods are derived with a bootstrap size of 100 repetitions. It is remarkable that the standard bootstrap method derives the confidence interval with a normality assumption, while the BCa bootstrap method introduces further bias and skewness corrections which weakens the reliance on normality. The numerical results in \cref{fig:confidence_interval_JP_0418_0501} reveal the consistency of confidence intervals among these methods. Despite the consistency, it is worth noting that bootstrap methods repeat the full computation multiple times which brings additional computational cost and problems when scaling up to business use cases. Conversely, our delta method based computation is one-pass and simultaneously yields confidence intervals with the evaluation of fairness metrics. This superiority renders our new method highly efficient and able to be deployed on large-scale complex business platforms.

By using bootstrap methods, we also estimate the bias of our fairness metric estimators. Our results are displayed in \cref{fig:bootstrap-bias}. It indicates that the biases of our estimators are lower than $3\%$. Furthermore, the oscillatory positive and negative sign of the bias suggests that there is no systematic bias in the estimation. These results are consistent with our theoretical analysis.

\subsection{Boosting Strategies and Metrics Changes}

In this subsection, we simulate treatment strategies data with artificially emulated \emph{boosting} strategies, which is equivalent to boosting the ranking 
score (used for ranking items in descending order and making the recommendations accordingly, when an item is boosted, its ranking goes up) of certain items. To simulate such strategies, we sample rows from daily recommendation logs with weights different for young adults and others. A larger weight stands for some boosting to the ranking 
score in the recommendation process. We note that the random traffic is still sampled uniformly without being affected by the boosting strategy. This is consistent with the definition of random traffic. We study three sets of boosting strategies where we give the mapping from the value of attribute young\_adult to the sampling weight as follows:

    (1) ``1.25x deboost'': \{0 : 1.25, 1 : 1\}, (2) ``2x deboost'': \{0 : 2, 1 : 1\}, (3) ``2x boost'': \{0 : 1, 1 : 2\}.

    \begin{wrapfigure}{l}{0.5\textwidth} 
    \includegraphics[width=.47\textwidth]{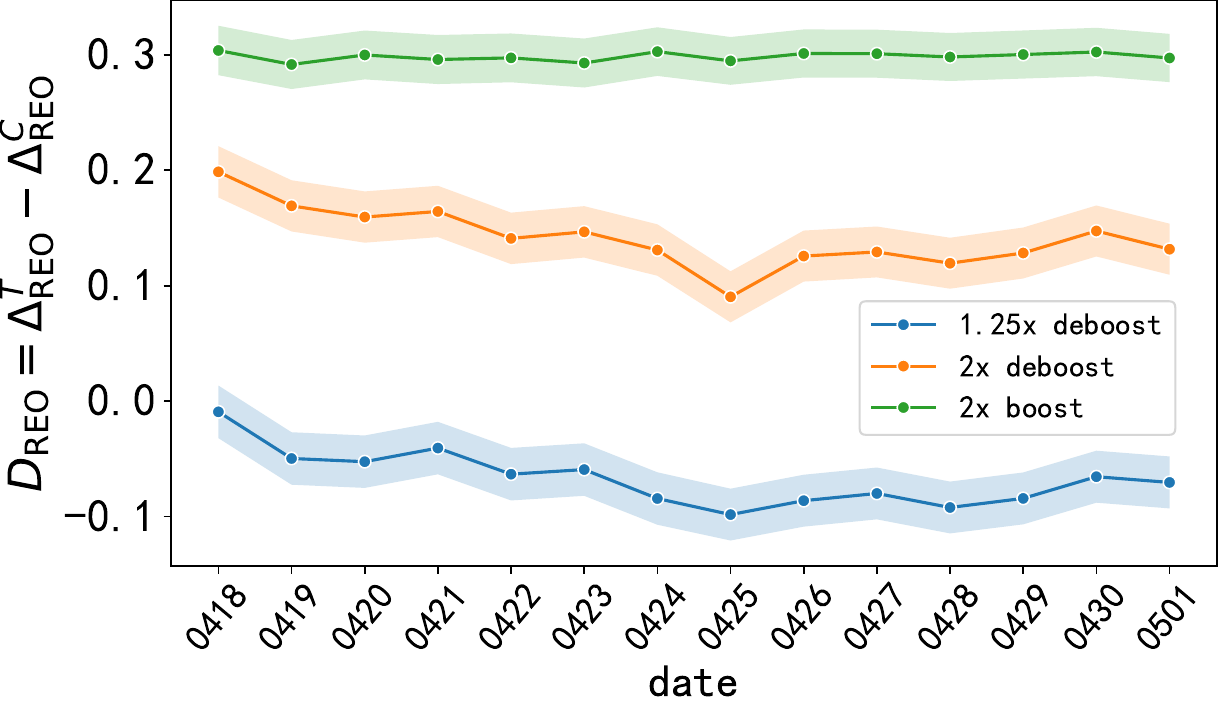}
        \caption{Monitoring difference-in-REO on simulated datasets with three boosting/deboosting strategies. }
        \label{fig:simulated-AB-test}
    \end{wrapfigure}
    
We monitor the difference in fairness penalties between the treatment group and the control group. The significance tests are derived by using \cref{alg:delta-method-AB}. In \cref{fig:simulated-AB-test}, we see that the strategy ``1.25x deboost'' can reduce the fairness penalty between young adults and others. This treatment strategy slightly deboosts the young-adult group, which is advantaged in the control strategy, and thus improves fairness. For more aggressive strategies, such as ``2x deboost'', the global fairness worsens, indicating an excessive suppression of the advantaged group and reverts the advantage. In the third strategy ``2x boost'', the advantaged group is further boosted which makes it even more advantaged. Consequently, the fairness penalty is further increased.

\section{Conclusion} \label{sec:conclusion}
In this work, we addressed the fairness metrics measurement in large-scale recommendation systems where preference labels are mostly missing. Specifically, for content creation platforms, fairness among creators and the users' personal preferences are equally important and REO is a popular metric that serves both purposes. To the best of our knowledge, we are the first to emphasize the necessity of including random traffic in production and data collection for accurate estimation of REO related metrics. We also designed efficient parametric inference techniques that leverage the random traffic and show both theoretically and numerically that it outperforms non-parametric methods like permutation tests in terms of efficiency and error tolerance. This efficient method serves as a novel recommendation fairness benchmarking tool. For label-dependent fairness measurement, we also proposed data collection methods for recommendation platforms, and provided a useful dataset from TikTok that is generated under this method. In the future, we aim to generalize our work to study the tradeoff between recommendation platforms' business goals and fairness measurement accuracy, as well as guidelines for choosing optimal random traffic volume for recommendation platforms.

\clearpage

\bibliographystyle{abbrvnat}

\clearpage
\appendix

\section{Related Works} \label{appsec:related}
Our work is closely related to the following research areas.

\paragraph{Fairness in Recommendation Systems} Fairness in recommendation systems \cite{Fair_Rec_Overview} studies the system's treatment disparities on different users \cite{User_Side_Fairness} and items \cite{REO_paper}. Compared with conventional fairness in machine learning studies \cite{Fair_ML_Review} on classification and regression models, fairness in recommendation systems measures the fairness of the entire system instead of single models, and better captures the system's unique nature like multi-sided properties \cite{Fair_Rec_Overview} and limited recommendation slots \cite{Fair_Exp_Rank, REO_paper}. The recommendation system for content creation platforms like YouTube, TikTok, and Twitter naturally consists of content creators and users (consumers), and fairness notions have been proposed on both the creator-side (item-side) \cite{REO_paper} and the user-side \cite{User_Side_Fairness}. Our work focuses on creator-side fairness where metrics like Fairness of Exposure \cite{Fair_Exp_Rank}, Ranking-based Equal Opportunity (REO), and Ranking-based Statistical Parity (RSP) \cite{REO_paper}, where REO is a metric that considers users' personal preferences and is one of the best fits for content creation platforms. To get an accurate measurement of these fairness metrics, we need ground truth information of every user-item pair in the dataset, where previous works run experiments on benchmark datasets like Yelp that satisfy this property. But in real-world large-scale recommendation systems, missing labels are prevalent since we can not obtain ground truth information on user-item pairs where the item was never recommended to the user.
We show that including random traffic is a necessity for measuring REO related metrics when having missing labels and elaborate on the measurement methods of other fairness metrics in the appendix, where either our method can generalize to their measurement or their measurement can be done without random traffic.

\paragraph{Random Traffic Data in Recommendation Systems}

Missing labels is a prevalent issue in large-scale recommendation systems, where the platform only has knowledge about users' preferences in the online traffic but has no knowledge about the massive missing interactions in the offline data \cite{KuaiRec}. It is well recognized that when models in the recommendation system only learn from previous online traffic, there exists exposure bias \cite{Chen_2023_TIS} that favors popular items, and such a bias will reinforce itself over time without further intervention. 
To tackle the exposure bias problem, various studies \cite{Yang_2018_RecSys} propose to use random traffic that recommends items uniformly at random, i.e., the missing at random (MAR) data \cite{Yang_2018_RecSys, Saito_2020_WSDM} to users to conduct exposure bias evaluations. Some companies also release their recommendation datasets with random traffic included, among which Yahoo releases popular public datasets \textbf{Yahoo!R3} \cite{Marlin_2009} and \textbf{Yahoo!R6} \cite{Li_2010_WWW}, and Kuaishou \cite{gao2022kuairand} releases a large random traffic dataset with over 300 million rows.

\section{Notation Table}

\begin{table}[!htb]
      \centering
        
        \begin{tabular}{p{0.2\textwidth} p{0.66\textwidth} }

        \toprule
        \textbf{Notation} & \textbf{Meanings} \\
        \midrule
        $u_m$ & The $m$-th user request
        \\

        \midrule
        $i_n$ & The $n$-th item
        \\

        \midrule
        $R(u_m, i_n)$ & The indicator of whether the default traffic recommended $i_n$ to request $u_m$
        \\

        \midrule
        $Y(u_m, i_n)$ & The indicator of user's preference on item $i_n$ for the user corresponding to request $u_m$
        \\

        \midrule
        $S(u_m, i_n)$ & The sensitive attribute of user-item (request-item) pair $(u_m, i_n)$
        \\

        \midrule
        $\mathcal{S} = \{s_1, \dots, s_K\}$ & The set of sensitive attributes
        \\
        
        \midrule
        $s_k$ & The $k$-th sensitive attribute value
        \\

        \midrule
        $\mathcal{R}_m$ & The set of recommended items given user request $u_m$
        \\

        \midrule
        $\mathcal{U}_{rand}, \mathcal{I}_{rand}, \mathcal{D}_{rand}$ & The set of user request, items and user-item pairs in the random traffic.
        \\

        \midrule
        $\stackrel{\cdot}{\sim}$ & Approximately distribute as.\\

        \midrule
        $\stackrel{a}{\sim}$ & Asymptotically distribute as.\\

        \midrule
        $\norm{U}_1$ & The 1-norm of the utility vector, which is equal to the sum of utilities $\norm{U}_1 = \sum_{j = 1}^K U_j$.\\

        \midrule
        $\norm{\cdot}_1, \norm{\cdot}_2, \norm{\cdot}_F$ & These notations stand for 1-norm, 2-norm and Frobenius norm respectively.\\
        
        \midrule
        $p_k$ & $p_k := P(Y(u, i) = 1, S(i) = s_k)$ which is the joint probability of an item liked by $u$ and the item from group $k$.\\

        \midrule
        $q_k$ & $q_k := P(Y(u, i) = 1, S(i) = s_k|R(u,i)=1)$ which is the joint probability of an item liked by $u$ and the item from group $k$ conditioned on that $i$ is recommended to $u$.\\

        \midrule
        $\hat{P}_k$ & The estimator of $p_k$.
        \\

        \midrule
        $\hat{Q}_k$ & The estimator of $q_k$.
        \\

        \midrule
        $U_k$ & The utility (Ranking-based true positive rate) of group $k$.
        \\

        \midrule
        $\hat{U}_k$ & The estimated utility of group $k$.
        \\

        \midrule
        $\Delta U_k$ & The relative utility of group $k$.
        \\

        \midrule
        $\wh{\Delta U_k}$ & The estimated relative utility of group $k$.
        \\

        \midrule
        $\Delta_{REO}$ & The fairness penalty.
        \\

        \midrule
        $\wh{\Delta_{REO}}$ & The estimated fairness penalty.
        \\
		\bottomrule
  
		\end{tabular}
  \vspace{2mm}
		\caption{Summary of notation.}\label{table:notation}
\end{table}

\section{Unidentifiability of REO}\label{app:identifiability}
The full dataset of user-item pairs is partitioned according to its recommendation or not $\Omega = \{R = 1\} \cup \{R = 0\}$. As users' preference is only accessible when they engage with the item, it implies that the preference label $Y$ is known on $\{R = 1\}$, while in contrast, on the unrecommended subset $\{R = 0\}$, the preference label $Y$ is missing. To capture the uncertainty of the unrecommended subset, we define the admissible set of unprobed dataset be a set of all full user-item pairs that agree with a given recommended subset.

\begin{definition}[Admissible set of unprobed dataset]
    Given a recommended subset $\Omega_1 = \{ \omega_j : R(\omega_j) = 1, \forall j = 1, \cdots, m_1\}$ and a integer $m_0$, an admissible unprobed dataset is a dataset consisting of $m_0 + m_1$ user-item pairs that agrees with $\Omega_1$ when restricting on the recommended subset $\Omega|_{R = 1} = \Omega_1$. Without further constraints, the admissible set of unprobed datasets is $\mathcal{F}_{\Omega_1, m_0} = \{\Omega_1 \cup \{\omega_j : R(\omega_j) = 0, Y(\omega_j) = y_j, S(\omega_j) = s_j, j = 1, \cdots, m_0 \} : y_j \in \{0, 1\}, s_j \in \mathcal{S} \forall j \}$ where the subset of unrecommended items exhausts all possible preference labels and sensitive attributes.
\end{definition}

Let $\phi(\Omega) \in \RR$ be a quantity to be computed on the dataset $\Omega$. In our case, it is the REO metric $\phi := \Delta_\mathrm{REO}$. Any algorithm produces a computational result $\phi_\mc{A}$ which is either exactly or approximately equal to the quantity of our interest, namely, the computation error $\abs{\phi(\Omega) - \phi_\mc{A}(\Omega)}$ is uniformly small for any dataset $\Omega$. For brevity, we do not consider randomized algorithm but our argument and analysis can be generalized to that case.

It is worth noting that when all recommended items corresponding to a sensitive attribute $s_k$ belong to the negative preference label $\{Y = 1, S = s_k\} \cap \Omega_1 = \emptyset$, the RTPR of that attribute $U_k = \Prob{R = 1 | Y = 1, S = s_k}$ is equal to zero by definition. Hence, when the recommended subset does not overlap with the positive preference of all sensitive attributes at all, the REO metric is unconditionally fair $\Delta_\mathrm{REO} = 0$. To isolate such trivial case, we consider a definition of nontrivial recommended subset as follows.
\begin{definition}[Nontrivial recommended subset]
    Let $\Omega_1$ be a subset of recommended user-item pairs. It is said to be nontrivial if there exists at least one pair with a positive preference label, namely, $\Omega_1 \cap \{Y = 1\} \ne \emptyset$.
\end{definition}

Note that the definition of nontrivial recommended subset rules out the case where all admissible unprobed dataset is unconditionally fair. In the proof of the following theorem, we show that there always exist a perfectly fair admissible dataset and a unfair admissible dataset for any nontrivial recommended subset. Consequently, any algorithm without probing the unrecommended subset can not compute these two cases accurately due to the lack of distinguishability.

\begin{lemma}\label{lem:exist-fair-and-unfair-sets}
    Given any nontrivial recommended subset $\Omega_1$ and a size parameter $m_0 \in \mathbb{N}_+$, there exist a perfectly fair admissible dataset $\Omega \in \mathcal{F}_{\Omega_1, m_0}$ and a unfair admissible dataset $\Omega^\prime \in \mathcal{F}_{\Omega_1, m_0}$.
\end{lemma}
\begin{proof}
      Note that the RTPR can be written as
      \begin{equation*}
          U_k=\frac{\mathbb{P}(R=1,Y=1,S=s_k)}{\mathbb{P}(R=1,Y=1,S=s_k)+\mathbb{P}(R=0,Y=1,S=s_k)}.
      \end{equation*}
      Because the recommended subset $\Omega_1$ is nontivial, there exists an index $k \in [K]$ so that $\mathbb{P}(R=1,Y=1,S=s_k) > 0$. Without loss of generality, we assume that it is the first sensitive attribute $s_1$. Otherwise, we may relabel subscripts to achieve it. Let the ratio be $\alpha := \mathbb{P}(R=0,Y=1,S=s_1) / \mathbb{P}(R=1,Y=1,S=s_1)$ which is well defined due to the positivity of the denominator. Furthermore, not that the denominator is fully fixed by the recommended subset $\Omega_1$ by definition. The ratio $\alpha$ is determined by the full dataset and $U_k = 1 / (1 + \alpha)$. 
      
      We consider the first admissible dataset $\Omega|_{R = 0} = \{(R = 0, Y = 0, S = s_1): i = 1, \cdots, m_0\}$. As the positive preference label only appears in the recommended subset, it concludes that $U_1 = \cdots = U_K = 1$ on $\Omega$ which implies a perfectly fair REO metric $\Delta_\mathrm{REO}(\Omega) = 0$.

      On the other hand, we can consider another admissible dataset $\Omega|_{R = 0} = \{(R = 0, Y = 1, S = s_1): i = 1, \cdots, m_0\}$ consisting of all positive preference corresponding to sensitive attribute $s_1$. Then, it gives a positive ratio $\alpha > 0$. Similar to the previous case, we have $U_2^\prime = \cdots = U_K^\prime = 1$ while $U_1^\prime = 1 / (1 + \alpha) < 1$ on $\Omega^\prime$. Direct computation gives
      \begin{equation*}
          \Delta U_1^\prime = \frac{K U_1^\prime}{U_1^\prime + \sum_{i = 2}^K U_i} - 1 = \frac{K}{1 + (1 + \alpha)(K-1)} - 1 = \frac{- \alpha (K - 1)}{1 + (1 + \alpha)(K-1)}.
      \end{equation*}
      Following symmetry $\Delta U_2^\prime = \cdots = \Delta U_K^\prime$ and conservation $\sum_{i = 1}^K \Delta U_i^\prime = 0$, it holds that
      \begin{equation*}
          \Delta U_2^\prime = - \frac{1}{K - 1} \Delta U_1^\prime = \frac{\alpha}{1 + (1 + \alpha) (K - 1)}.
      \end{equation*}
      Hence, the REO metric on $\Omega^\prime$ is
      \begin{equation*}
        \Delta_\mathrm{REO}(\Omega^\prime) = \frac{\alpha K \sqrt{K - 1}}{1 + (1 + \alpha) (K - 1)} \ne \Delta_\mathrm{REO}(\Omega) = 0.
      \end{equation*}
\end{proof}

Using the proceeding lemma, we can prove the unidentifiability of REO metric.

\begin{theorem}[formal version of \cref{thm:identifiability}]\label{thm:identifiability-rigorous}
    Given any nontrivial recommended subset $\Omega_1$ and a size parameter $m_0 \in \mathbb{N}_+$, there exist two admissible datasets $\Omega, \Omega^\prime \in \mathcal{F}_{\Omega_1, m_0}$ so that their REO metrics cannot be simultaneously computed to arbitrary precision by any algorithm without probing the unrecommended subset. That means that there is an $\epsilon_0 > 0$ so that for any algorithm $\mc{A}$ without probing the unrecommended subset, either $|\phi(\Omega) - \phi_\mathcal{A}(\Omega)| \ge \epsilon_0$ or $|\phi(\Omega^\prime) - \phi_\mathcal{A}(\Omega^\prime)| \ge \epsilon_0$ holds. 
\end{theorem}

\begin{proof}
    Without probing the unrecommended subset, the algorithm can not distinguish two admissible datasets $\Omega, \Omega^\prime \in \mathcal{F}_{\Omega_1, m_0}$, namely, $\phi_\mathcal{A}(\Omega)=\phi_\mathcal{A}(\Omega^\prime)$. Applying triangle inequality, it holds that 
    \begin{equation*}
        \begin{split}
            |\phi(\Omega)-\phi(\Omega^\prime)|&=|\phi(\Omega)-\phi_\mathcal{A}(\Omega)+\phi_\mathcal{A}(\Omega^\prime)-\phi(\Omega^\prime)|\\
            &\le|\phi(\Omega)-\phi_\mathcal{A}(\Omega)|+|\phi(\Omega^\prime)-\phi_\mathcal{A}(\Omega^\prime)|.
        \end{split}
    \end{equation*}
    Let $\epsilon_0:=|\phi(\Omega)-\phi(\Omega^\prime)|/2$ which is strictly positive according to \cref{lem:exist-fair-and-unfair-sets}. If the computation errors of both datasets are less than $\epsilon_0$, it contradicts the derived inequality. Consequently, these two datasets can not be computed accurately at the same time. 
\end{proof}

As a closing remark of this section, there might be other weighted sampling strategies to be adopted rather than uniformaly random traffic. Such change-of-weight can be understood in the framework of importance sampling where an importance weight is adjusted to ensure unbiased estimation. Let $\pi$ be a sampling procedure defined as a probability distribution on the full set of user-item pairs consisting of $m_0$ elements. Then, the unbiased estimation for any quantity $A$ is $\mathbb{E}(A) = \mathbb{E}_{\omega \sim \pi}(A(\omega) / \pi(\omega))$ where user-item pair $\omega \sim \pi$ is drawn from the changed distribution. Note that exactly tracking the importance weight $1 / \pi(\omega)$ becomes increasingly challenging when the subset size $m_0$ becomes large which is common in practice. Consequently, using a nonuniform sampling strategy might produce unwanted bias and error alongside the computation of REO. This argument suggests the superiority of using random traffic over using other nonuniform sampling strategies. 

\section{Fairness Metrics and Fairness Monitoring}\label{sec:fairness-monitoring}

In this part, we discuss the actual application of REO fairness monitoring. We will discuss the metrics we measure within a recommendation strategy (which can apply to global fairness estimation) and between different recommendation strategies (which can apply to A/B tests). Specifically, a recommendation strategy corresponds to a certain system configuration, including the model, filtering rules, boosting rules, UI design, etc.

\begin{remark}
    A/B tests can concurrently occur on the platform with a very high number of distinct configurations (see \cref{sec:discussion} for elaboration). A strategy can be a high-level concept, which is a combination of many fine-grained strategies. For a simplified example where the system contains two configuration dimensions, one for model and the other for UI, the most fine-grained configuration can contain the Cartesian product of (use model A, use model B) $\times$ (use UI C, use UI D), corresponding to 4 most fine-grained strategies. But the strategies can also be higher-level concepts like ``use model A'' vs ``use model B'' (while in each strategy the UI design distribution is the same). 
\end{remark}

Note that the definition of the group utility in \cref{eq:REO_utility} involves counterfactual events as the users' preference label is identified after they receive the recommended item. The causality renders the direct evaluation of the group utility infeasible in accordance with the identifiability argument in the previous subsection. To compute REO related metrics, we need to factor the group utility into measurable components by leveraging default and random traffic. Using Bayesian theorem, we have 
\begin{equation*}
    U_k = \Prob{R = 1 | Y = 1, S = s_k} = \frac{\Prob{Y = 1, S = s_k | R = 1}}{\Prob{Y = 1, S = s_k}} \Prob{R = 1}.
\end{equation*}
Then, the group utility is assembled from three probabilities, which are either measurable or irrelevant as discussed in the paragraph below.

For notational simplicity, we denote $p_k := P(Y = 1, S = s_k)$ and $q_k := P(Y = 1, S = s_k | R = 1)$ which are frequently used in the analysis in the rest of this work. At a high level, $p_k$ measures the proportion of ``preferred samples'' from group $k$ in the entire candidate pool while $q_k$ measures such proportions in the recommended candidates. Ideally, for fair systems in terms of REO, we should observe similar values between $p_k$ and $q_k$ for all $k$. We also define the following 
\begin{equation}
    \hat{P}_{k} := \frac{\sum_{(u, i) \in \mathcal{D}_{rand}} \mathbb{I}(Y(u, i)=1) \mathbb{I}(S(i) = s_k)}{|\mathcal{D}_{rand}|}, \hat{Q}_{k} := \frac{\sum_{(u, i) \in \mathcal{D}_{rec}} \mathbb{I}(Y(u, i)=1) \mathbb{I}(S(i) = s_k)}{|\mathcal{D}_{rec}|}
\end{equation}
as the noisy estimators of $p_k$ and $q_k$ from $\mathcal{D}_{rand}$ and $\mathcal{D}_{rec}$. Then we define
\begin{equation}
    \hat{U}_k := \frac{\hat{Q}_k}{\hat{P}_k},~~~~ \wh{\Delta_\mathrm{REO}} := \frac{\mathrm{std}(\hat{U}_1, \dots, \hat{U}_K)}{\mathrm{mean}(\hat{U}_1, \dots, \hat{U}_K)}.
\end{equation}
It is worth noting that the estimator $\hat{U}_k$ is equal to the the utility function up to a scalar $\Prob{R = 1}$ which is independent with the group index ($k$). Note that the fairness penalty is invariant under simultaneous rescaling of utility functions. Hence, this difference in the estimator will not affect the estimation of the metric of interest. The rescaling scalar $\Prob{R = 1}$ stands for the probability that a random user-item pair goes through a recommendation process. Due to the extremely large size of the possible user-item pairs and the large volume of the recommendation log, exactly measuring this quantity is highly infeasible. However, its appearance in the utility function is inevitable due to the counterfactual nature. Thanks to Bayesian theorem, we factor the utility function as two measurable quantities using default and random traffic, and we get rid of the unmeasurable probability according to the scale invariance of fairness penalty.

\paragraph{Within a Strategy.} Apart from fairness penalty, relative group utilities $\Delta U_k$'s are metrics indicating the advantage or disadvantage of groups. In actual production, we use the following for estimation
\begin{equation}
    \wh{\Delta U_k} := \frac{\hat{U}_k}{\mathrm{mean}(\hat{U}_1,\dots,\hat{U}_K)} - 1.
\end{equation}
Due to the concentration of our related scalar metrics in high dimensional data space, fairness metrics can be estimated accurately with uniformly sampled data from the full log data. To quantify the accuracy of the estimation, we present the following theorem. A series of more comprehensive statements and the proof are presented in \cref{sec:proofs}.

\begin{theorem}\label{thm:main_estimate__reo_metrics}
    Assume that the joint portion of each group in the traffic is a nonvanishing constant, namely $p_k, q_k = \Omega(1)$ for any $k = 1, \cdots, K$. Suppose the sample size is $\abs{\mathcal{D}_{rec}}, \abs{\mathcal{D}_{rand}} = O\left( K^2 \epsilon^{-2} \log\left(K \delta^{-1}\right)\right)$, then with probability at least $1 - \delta$, the estimation errors are uniformly upper bounded:
    \begin{equation*}
        \max_{k = 1, \cdots, K} \abs{\wh{\Delta U_k} - \Delta U_k} \le \epsilon \text{ and } \abs{\wh{\Delta_\mathrm{REO}} - \Delta_\mathrm{REO}} \le \epsilon.
    \end{equation*}
\end{theorem}

It is worth noting that $p_k, q_k$ are estimated from sample means which are asymptotically normal distributed following central limit theorem. Intuitively, as a function transformation of them, our fairness metrics should also admit a similar asymptotic normality. This intuition is quantified in the analysis and proof presented in \cref{sec:proofs}. Consequently, using delta method \cite{Doob1935}, the variance of fairness metrics can be derived by propagating the variance of sources $p_k, q_k$'s. This gives arise to an extremely simple method for quantifying the confidence intervals of the estimators of fairness metrics which is outlined in \cref{alg:delta-method}.

\begin{algorithm}[htbp]
\caption{Computing fairness metrics, their standard errors and confidence intervals}
\label{alg:delta-method}
\begin{algorithmic}
\State \textbf{Input: } Default traffic $\mathcal{D}_{rec}$ and random traffic $\mathcal{D}_{rand}$ with $Y$, $S$, and confidence level $1 - \delta$.
\State Compute sample means $\hat{P}_k = \mathbb{P}(Y(u, i) = 1, S(i) = s_k)$ on $\mathcal{D}_{rand}$. 
\State Compute sample means $\hat{Q}_k = \mathbb{P}(Y(u, i) = 1, S(i) = s_k | R(u, i) = 1)$ on $\mathcal{D}_{rec}$.
\State Compute the group utility estimators $\hat{U}_k = \hat{Q}_k / \hat{P}_k$ of each group $k = 1, \cdots, K$.
\State Compute the REO penalty estimator $\wh{\Delta_\mathrm{REO}}$ from $\{\hat{U}_k\}_{k=1}^K$.
\State Compute the relative group utility function $\wh{\Delta U_k}$ of each group $k = 1, \cdots, K$.
\State {\textcolor{gray}{// Deriving confidence intervals}}
\State Assemble a $K$-by-$K$ diagonal matrix $\Gamma_{k, k} = \hat{U}_k^2 \left(\frac{1 - \hat{Q}_k}{\hat{Q}_k} \frac{1}{\abs{\mc{D}_{rec}}} + \frac{1 - \hat{P}_k}{\hat{P}_k} \frac{1}{\abs{\mc{D}_{rand}}} \right)$. {\textcolor{gray}{// variance of modified group utility functions}}
\State Assemble a $K$-by-$K$ matrix $G_{j, k} = K \frac{\delta_{j, k} \sum_{i = 1}^K \hat{U}_i - \hat{U}_k}{(\sum_{i = 1}^K \hat{U}_i)^2}$. {\textcolor{gray}{// Jacobian matrix of the transformation $(U_1, \cdots, U_K) \mapsto (\Delta U_1, \cdots, \Delta U_K)$}}
\State Assemble a $K$-dimensional vector $H_j = \frac{\wh{\Delta U_j}}{K \wh{\Delta_\mathrm{REO}}}$. {\textcolor{gray}{// Gradient of the transformation $(\Delta U_1, \cdots, \Delta U_K) \mapsto \Delta_\mathrm{REO}$}}
\State Set $\mathrm{se}(\wh{\Delta_\mathrm{REO}}) \leftarrow \mathrm{std}(\wh{\Delta_\mathrm{REO}}(\cdot))$.
\State Compute the covariance matrix $[\mathrm{Cov}(\wh{\Delta U_j}, \wh{\Delta U_k})]_{j, k} = \Sigma = G^\top \Gamma G$, and the variance $\mathrm{Var}(\wh{\Delta_\mathrm{REO}}) = \Xi = H^\top \Sigma H$. 
\State Set $z(\delta / 2)$ to the upper quantile of the standard normal distribution.
\For{$k = 1, \cdots, K$}
\State For $\Delta U_k$, set the standard error $\mathrm{SE}(\wh{\Delta U_k}) = \sqrt{\Sigma_{k, k}}$ and the confidence interval $\wh{\Delta U_k} \pm z(\delta / 2) \mathrm{SE}(\wh{\Delta U_k})$.
\EndFor
\State For $\Delta_\mathrm{REO}$, set the standard error $\mathrm{SE}(\wh{\Delta_\mathrm{REO}}) = \sqrt{\Xi}$ and the confidence interval $\wh{\Delta_\mathrm{REO}} \pm z(\delta / 2) \mathrm{SE}(\wh{\Delta_\mathrm{REO}})$.
\State \textbf{Output: } Estimators, standard errors, and confidence intervals.
\end{algorithmic}
\end{algorithm}

\paragraph{Between Strategies.} A/B testing is the most popular approach to test if an alternative strategy is superior to the current strategy, e.g., replacing a recommendation model with a more sophisticated one. To the best of our knowledge, while most platforms have core business metrics' statistical significance analyses in A/B tests, there isn't much discussion on fairness metrics' statistical significance, where the latter is important for building responsible and sustainable platforms. We want to make sure: (1) the new strategy will not significantly worsen the system fairness, (2) fairness enhancing strategies have significant effects, or (3) new strategies are not blocked by false positive fairness alarms.

For clarity of presentation, we will show our results in A/B testing cases with a control group and a treatment group, but our analysis and methods generalize to an arbitrary number of experiment groups. We denote the utilities in the control and treatment group as $U_k^C, U_k^T, \hat{U}_k^C, \hat{U}_k^T$, which are measured on $\mathcal{D}_{rec}^C, \mathcal{D}_{rec}^T$, with a common $\mathcal{D}_{rand}$. We will measure if the treatment strategy has a statistically significant REO fairness metric change by measuring the following metrics:
\begin{enumerate}
    \item For any group index $k = 1, \cdots, K$, $\wh{D_k} := \wh{\Delta U_k^T} - \wh{\Delta U_k^C}$ measures the difference. Monitoring it ensures that the disadvantaged groups in the control strategy will not be further disadvantaged in the treatment group.
    \item $\wh{D_\mathrm{REO}} := \wh{\Delta_\mathrm{REO}^T} - \wh{\Delta_\mathrm{REO}^C}$ measures the change in the global fairness penalty. Monitoring it ensures that the global fairness penalty does not grow significantly in the treatment strategy.
\end{enumerate}

According to the construction, the fairness metrics in control and treatment groups in the A/B test data share a common random traffic $\mathcal{D}_{rand}$. It might introduce unwanted correlations between estimators from two groups. However, notice that $\hat{P}_k$ is increasingly close to a scalar constant $p_k$ when the size of the random traffic $\abs{\mathcal{D}_{rand}}$ is large enough according to the law of large numbers. Then, it is reasonable to assume that the correlation between fairness metric estimators from control and treatment groups are weak in the large sample limit in terms of $\abs{\mathcal{D}_{rand}}$. Using the testing method in \cref{alg:delta-method}, the significance tests of fairness metrics in A/B experiments can be established similarly. We outline the method in \cref{alg:delta-method-AB}.

\begin{algorithm}[htbp]
\caption{Testing fairness metrics in A/B experiments}
\label{alg:delta-method-AB}
\begin{algorithmic}
\State \textbf{Input: } Default traffic $\mathcal{D}_{rec}^C$ of control group, default traffic $\mathcal{D}_{rec}^T$ of treatment group and random traffic $\mathcal{D}_{rand}$ with fields preference label $Y$ and sensitive attribute $S$, confidence level $1 - \delta$.
\State Compute estimators $\wh{\Delta_\mathrm{REO}^C}$, $\wh{\Delta U_k^C}, k = 1, \cdots, K$ and their standard errors using $\mathcal{D}_{rec}^C$, $\mathcal{D}_{rand}$ and \cref{alg:delta-method}.
\State Compute estimators $\wh{\Delta_\mathrm{REO}^T}$, $\wh{\Delta U_k^T}, k = 1, \cdots, K$ and their standard errors using $\mathcal{D}_{rec}^T$, $\mathcal{D}_{rand}$ and \cref{alg:delta-method}.
\State Set $z(\delta / 2)$ to the upper quantile of the standard normal distribution.
\For{$k = 1, \cdots, K$}
\State Set $\wh{D_k} = \wh{\Delta U_k^T} - \wh{\Delta U_k^C}$, set the standard error $\mathrm{SE}(\wh{D_k}) = \sqrt{\mathrm{SE}^2(\wh{\Delta U_k^T}) + \mathrm{SE}^2(\wh{\Delta U_k^C})}$ and the confidence interval $\wh{D_k} \pm z(\delta / 2) \mathrm{SE}(\wh{D_k})$.
\EndFor
\State Set $\wh{D_\mathrm{REO}} = \wh{\Delta_\mathrm{REO}^T} - \wh{\Delta_\mathrm{REO}^C}$, set the standard error $\mathrm{SE}(\wh{D_\mathrm{REO}}) = \sqrt{\mathrm{SE}^2(\wh{\Delta_\mathrm{REO}^T}) + \mathrm{SE}^2(\wh{\Delta_\mathrm{REO}^C})}$ and the confidence interval $\wh{D_\mathrm{REO}} \pm z(\delta / 2) \mathrm{SE}(\wh{D_\mathrm{REO}})$.
\State \textbf{Output: } Estimators, standard errors, and confidence intervals.
\end{algorithmic}
\end{algorithm}

\section{Deferred proofs}\label{sec:proofs}
\begin{theorem} \label{thm:estimate-pk}
    $\hat{P}_k$ and $\hat{Q}_k$ are unbiased estimators of $p_k$ and $q_k$ respectively. Furthermore, it suffices to set the random traffic size to $\abs{\mathcal{D}_{rand}} = O(p_k^{-1} \epsilon^{-2} \log(\delta^{-1}))$ and the default traffic size to $\abs{\mathcal{D}_{rec}} = O(q_k^{-1} \epsilon^{-2} \log(\delta^{-1}))$ so that both estimators is $\epsilon$-close to the exact value in relative error with probability at least $1 - \delta$.
\end{theorem}
\begin{proof}[Proof of \cref{thm:estimate-pk}]
    Because the user request $u_m$ and item $i_n$ are chosen independently, the summand in the numerator of $\hat{P}_{k}$ consists of iid Bernoulli random variables. Hence, according to Chernoff bound, it holds that
    \begin{equation*}
        \Prob{\abs{\hat{P}_{k} - p_k} / p_k \ge \epsilon} \le 2 e^{- \abs{\mathcal{D}_{rand}} \epsilon^2 p_k / 3}.
    \end{equation*}
    Choosing $\abs{\mathcal{D}_{rand}} = \lceil 3 p_k^{-1} \epsilon^{-2} \log(2 \delta^{-1})\rceil$, the above probability is bounded by $\delta$, which proves the theorem. Similarly, the result can be derived for $\hat{Q}_k$ on default traffic.
\end{proof}

As multiple quantities with distinct subscript $k$ are estimated collectively, it is convenient to extract a sample-size control parameter $n$ from sizes of all datasets. The following analysis suggests
\begin{equation}\label{eqn:sample-size}
    \abs{\mathcal{D}_{rec}} = n \max_k U_k^2 \frac{1 - q_k}{q_k} \text{ and } \abs{\mathcal{D}_{rand}} = n \max_k U_k^2 \frac{1 - p_k}{p_k}
\end{equation}
which largely simplify the presentation of the theoretical results. The parameter $n$ is referred to as the sample size parameter, which controls the size of both datasets simultaneously.

The estimation performance of group utilities can be derived similarly.

\begin{theorem}\label{thm:estimate-Uk}
    The estimator $\hat{U}_k$ is a consistent estimator of the RTPR utility. When the sample size parameter is $n = O(\epsilon^{-2} \log(\delta^{-1}))$, the estimator (up to a multiplicative constant) is $\epsilon$-close to the RTPR utility with probability at least $1 - \delta$.
\end{theorem}
\begin{proof}[Proof of \cref{thm:estimate-Uk}]
    Note that in both $\hat{P}_k$ and $\hat{Q}_k$, the summand consists of iid samples, the defined quantity converges in probability to the exact values $\hat{P}_k \stackrel{p}{\to} p_k$ and $\hat{Q}_k \stackrel{p}{\to} q_k$ as the traffic sizes are large enough. As a consequence of Slutsky's theorem, the quotient $\hat{U}_k$ converges in probability to the quantity $U_k$ up to a scaling constant $C := \Prob{R = 1}$ which is independent with $k$. According to central limit theorem, the numerator and denominator are approximately normal distributed:
    \begin{equation*}
        \hat{Q}_k \stackrel{a}{\sim} N(q_k, q_k(1 - q_k) / \abs{\mc{D}_{rec}}) \text{ and } \hat{P}_{ k} \stackrel{a}{\sim} N(p_k, p_k(1 - p_k) / \abs{\mc{D}_{rand}}).
    \end{equation*}
    According to \cite{DiazEloisa2013}, when $\abs{\mc{D}_{rec}}, \abs{\mc{D}_{rand}} \gg 1$ are large enough, the quotient is well approximated by a normal distribution:
    \begin{equation*}
        \frac{\hat{Q}_k}{\hat{P}_{ k}} \stackrel{\cdot}{\sim} N\left(\frac{q_k}{p_k}, \frac{q_k^2}{p_k^2} \left(\frac{1 - q_k}{q_k} \frac{1}{\abs{\mc{D}_{rec}}} + \frac{1 - p_k}{p_k} \frac{1}{\abs{\mc{D}_{rand}}} \right)\right).
    \end{equation*}
    Consequently, plugging the scaling constant into the estimator, the above approximation implies that
    \begin{equation}
    \begin{split}
        \frac{\hat{U}_k}{C} & \stackrel{\cdot}{\sim} N\left(U_k, U_k^2 \left(\frac{1 - q_k}{q_k} \frac{1}{\abs{\mc{D}_{rec}}} + \frac{1 - p_k}{p_k} \frac{1}{\abs{\mc{D}_{rand}}} \right)\right) = N\left(U_k, \frac{2}{n}\right)
    \end{split}
    \end{equation}
    where the definition of sample sizes is used.
    Hence
    \begin{align*}
        \Prob{\abs{\hat{U}_k / C - U_k} \ge \tilde{\epsilon} \sqrt{2 / n}} = \frac{2}{\sqrt{\pi}} \int_{\tilde{\epsilon} / \sqrt{2}}^\infty e^{- t^2} \ud t = \mathrm{erfc}(\tilde{\epsilon} / \sqrt{2}) \le e^{- \tilde{\epsilon}^2 / 2}.
    \end{align*}
    It suffices to choose $n = \lceil 4 \epsilon^{-2} \log(\delta^{-1}) \rceil$ so that the estimation error is upper bounded by $\epsilon$ with probability at least $1 - \delta$.
\end{proof}

\begin{theorem}\label{thm:estimate-delta-Uk}
    The estimator $\wh{\Delta U_k}$ is a consistent estimator of the relative group utility. When the sample size parameter is $n = O\left(\frac{K^2}{\norm{U}_1^2 \epsilon^2} \log\left(\frac{K}{\delta}\right)\right)$, the estimation error is uniformly upper bounded:
    \begin{equation*}
        \max_{k = 1, \cdots, K} \abs{\wh{\Delta U_k} - \Delta U_k} \le \epsilon
    \end{equation*}
    with probability at least $1 - \delta$.
\end{theorem}

\begin{proof}[Proof of \cref{thm:estimate-delta-Uk}]
    The consistency follows continuous mapping theorem. Note that the estimator $\wh{\Delta U_k}$ is invariant by rescaling $\hat{U}_k$ by any constant independent with the subscript index $k$. Let $\tilde{U}_k := \hat{U}_k / C$ be the estimator of $U_k$ whose distribution is derived in the proof of \cref{thm:estimate-Uk}. Then, $\wh{\Delta U_k}$ can be derived from $\tilde{U}_k$ with the same formula. Let us consider a $K$-variate function $g : \RR^K \to \RR^K$ which is component-wisely defined as
    \begin{equation}
        g_k(U_1, \cdots, U_K) = K \frac{U_k}{\sum_{i=1}^K U_i} - 1.
    \end{equation}
    The Jacobian of the function is element-wisely defined as
    \begin{equation*}
        \partial_j g_k := \frac{\partial g_k}{\partial U_j} = K \frac{\delta_{k, j} \sum_{i = 1}^K U_i - U_k}{(\sum_{i = 1}^K U_i)^2}
    \end{equation*}
    where $\delta_{k, i}$ is the Kronecker delta. The proof of \cref{thm:estimate-Uk} indicates that when the numbers of samples are chosen as 
    \begin{equation*}
        \abs{\mc{D}_{rec}} = n  \max_k U_k^2 \frac{1 - q_k}{q_k} \text{ and } \abs{\mc{D}_{rand}} = n  \max_k U_k^2 \frac{1 - p_k}{p_k}
    \end{equation*}
    and when $n$ is sufficiently large, estimators $\wt{U}_k$'s are asymptotically jointly normal distributed as
    \begin{equation*}
        \sqrt{n} (\wt{U}_k - U_k) \to N(0, \varsigma_k^2) \text{ with } \varsigma_k^2 \le 2.
    \end{equation*}
    Applying delta method \cite{Doob1935}, it holds that
    \begin{equation}
        \sqrt{n} (g(\wt{U}_1, \cdots, \wt{U}_K) - g(U_1, \cdots, U_K)) \to N(0, \Sigma).
    \end{equation}
    Here, $\Sigma$ is covariance matrix element-wisely bounded as
    \begin{equation}
        \abs{\Sigma_{k, l}} \le 2 \abs{\sum_{k = 1}^K \partial_j g_k \partial_j g_l}.
    \end{equation}
    To study the sample complexity of estimating $\hat{\Delta} U_k = g_k(\wt{U}_1, \cdots, \wt{U}_K)$, it suffices to upper bound the diagonal element of the covariance
    \begin{equation}\label{eqn:diagonal-covariance-bound}
    \begin{split}
        \abs{\Sigma_{k, k}} &\le 2 \sum_{j = 1}^K \abs{\partial_j g_k}^2 = 2 K^2  \frac{\left(\sum_{k \ne j} U_k \right)^2 +  \sum_{k \ne j} U_k^2}{(\sum_{i = 1}^K U_i)^4}\\
        &\le 2 K^2 \frac{2 \left(\sum_{k \ne j} U_k \right)^2}{(\sum_{i = 1}^K U_i)^4} \le \frac{4  K^2}{(\sum_{i = 1}^K U_i)^2}.
    \end{split}
    \end{equation}
    Hence, to achieve $\max_{k = 1, \cdots, K} \abs{\hat{\Delta} U_k - \Delta U_k} \le \epsilon$ with probability at least $1 - \delta$, following union bound, it suffices to choose
    \begin{equation}
        n = \frac{4 K^2}{(\sum_{i = 1}^K U_i)^2 \epsilon^2} \log(K / \delta) = O\left(\frac{K^2}{\norm{U}_1^2 \epsilon^2} \log\left(\frac{K}{\delta}\right)\right).
    \end{equation}
\end{proof}

\begin{theorem}\label{thm:estimate-REO}
    $\wh{\Delta_\mathrm{REO}}$ is a consistent estimator of fairness penalty. Furthermore, when the sample size parameter $n = O\left(\frac{K^2}{\norm{U}_1^2 \epsilon^2} \log\left(\frac{1}{\delta}\right)\right)$, the estimator is $\epsilon$-close to fairness penalty with probability at least $1 - \delta$.
\end{theorem}

\begin{proof}[Proof of \cref{thm:estimate-REO}]
    The consistency follows continuous mapping theorem. Note that when $j \ne l$, it holds that
    \begin{equation*}
    \begin{split}
        \abs{\Sigma_{k, l}} &\le 2 \abs{\sum_{j = 1}^K \partial_j g_k \partial_j g_l}\\
        &= \frac{2 K^2}{(\sum_{i = 1}^K U_i)^4} \abs{ K U_k U_l - (U_k + U_l) \sum_{i = 1}^K U_i }.
    \end{split}
    \end{equation*}
    Then, we have
    \begin{equation}
        \begin{split}
            &\sum_{k \ne l} \abs{\frac{(\sum_{i = 1}^K U_i)^4}{2 K^2} \Sigma_{k, l}}^2 \le \sum_{k \ne l} \left( K^2 U_k^2 U_l^2 + (U_k + U_l)^2 (\sum_i U_i)^2 \right)\\
            & \le \sum_{k = 1}^K \sum_{l = 1}^K \left( K^2 U_k^2 U_l^2 + (U_k + U_l)^2 (\sum_i U_i)^2 \right)\\
            &\le K^2 (\sum_i U_i^2)^2 + 2 (\sum_i U_i^2) (\sum_i U_i)^2 + 2 (\sum_i U_i)^4\\
            &\le (K^2 + 4) (\sum_i U_i)^4.
        \end{split}
    \end{equation}
    Meanwhile, according to \cref{eqn:diagonal-covariance-bound}, the diagonal elements of the covariance matrix satisfy
    \begin{equation}
        \sum_{k = 1}^K \abs{\frac{(\sum_{i = 1}^K U_i)^4}{2 K^2} \Sigma_{k, k}}^2 \le 4 K (\sum_i U_i)^4. 
    \end{equation}
    Consequently, the sum of square of the matrix elements of the covariance matrix, which is equal to the squared Frobenius norm of the covariance matrix, is upper bounded as
    \begin{equation}
    \begin{split}
        \norm{\Sigma}_F^2 &= \sum_{k = 1}^K \sum_{l = 1}^K \abs{\Sigma_{k, l}}^2 \le \left(\frac{2 K^2}{(\sum_i U_i)^4}\right)^2 (K^2 + 4K + 4) (\sum_i U_i)^4\\
        & = \frac{4 K^4(K+2)^2}{(\sum_i U_i)^4} \le \frac{16 K^6}{(\sum_i U_i)^4}.
    \end{split}
    \end{equation}
    Here, the last inequality uses the fact that $K \ge 2$.
    Hence, the 2-norm of the covariance matrix is upper bounded as
    \begin{equation}
        \norm{\Sigma}_2 \le \norm{\Sigma}_F \le \frac{4   K^3}{(\sum_{i = 1}^K U_i)^2}.
    \end{equation}
    Let $h : \RR^K \to \RR$ be a function defining the formula for computing REO
    \begin{equation}
        h(\Delta U_1, \cdots, \Delta U_K) = \sqrt{\frac{1}{K} \sum_{i = 1}^K (\Delta U_i)^2}.
    \end{equation}
    It is straightforward to show that this function coincides with the REO in the sense that $h(\Delta U_1, \cdots, \Delta U_K) = \Delta_\mathrm{REO}$. Note that a use $1 / K$ in the expression of $\mathrm{std}(\cdots)$ for simplicity. When switching to $1 / (K - 1)$, the complexity scaling remains the same. The partial derivative of ths function is
    \begin{equation*}
        \partial_j h := \frac{\partial h}{\partial (\Delta U_j)} = \Delta U_j \sqrt{\frac{1}{K \sum_{i = 1}^K (\Delta U_i)^2}}.
    \end{equation*}
    Following delta method, the asymptotic distribution of the REO estimator is
    \begin{equation}
        \sqrt{n} (\wh{\Delta_\mathrm{REO}} - \Delta_\mathrm{REO}) \to N(0, \Xi). 
    \end{equation}
    Here, the variance is
    \begin{equation*}
        \Xi = \sum_{j, l} \partial_j h \partial_l h \Sigma_{j, l} \le \norm{\Sigma}_2 \sum_{j = 1}^K (\partial_j h)^2.
    \end{equation*}
    Note that
    \begin{equation*}
        \sum_{j = 1}^K (\partial_j h)^2 = \frac{1}{K \sum_{i = 1}^K (\Delta U_i)^2} \sum_{j = 1}^K (\Delta U_j)^2 = \frac{1}{K}.
    \end{equation*}
    Thus, the variance is upper bounded as
    \begin{equation*}
        \Xi \le \frac{4   K^2}{(\sum_{i = 1}^K U_i)^2}.
    \end{equation*}
    Consequently, to estimate REO fairness penalty up to error $\epsilon$ and with probability at least $1 - \delta$, it suffices to choose
    \begin{equation}\label{eqn:estimate-REO-n-size}
        n = \frac{4   K^2}{(\sum_{i = 1}^K U_i)^2 \epsilon^2} \log(1 / \delta) = O\left(\frac{K^2}{\norm{U}_1^2 \epsilon^2} \log\left(\frac{1}{\delta}\right)\right).
    \end{equation}
\end{proof}

\begin{remark}
    Before closing the section of proofs, we remark on the sample complexity requirement on datasets for accurate estimation results. To estimate relative group utilities and fairness penalty simultaneously accurate, we can take the maximum among sample complexities given in \cref{thm:estimate-delta-Uk,thm:estimate-REO}, which requires $n = O(K^2 \norm{U}_1^{-2} \epsilon^{-2} \log(K \delta^{-1}))$. When proportions $p_k, q_k$ are constantly large, the relation between the parameter $n$ and traffic sizes in \cref{eqn:sample-size} can be simplified. For example, the following holds
    \begin{equation*}
        O(n \max_k U_k^2) = O\left( \frac{K^2 \max_k U_k^2}{\norm{U}_1^2 \epsilon^2} \log(K / \delta) \right) \le O(K^2 \epsilon^{-2} \log(K / \delta))
    \end{equation*}
    where $\norm{U}_1 \ge \max_k U_k$ is used. Hence, it suffices to set the size of default traffic to $O(K^2 \epsilon^{-2} \log(K / \delta))$ to ensure an accurate estimation result. Similarly, the size of random traffic is sufficient to be set to $O(K^2 \epsilon^{-2} \log(K / \delta))$ for an accurate estimation. This proves \cref{thm:main_theorem_estimation}.
\end{remark}

\section{Justifying \cref{alg:delta-method}}
In this section, we justify \cref{alg:delta-method} in a theoretical manner by going through the analysis using delta method.

    Let $H := \nabla h$ be the gradient vector which is element-wisely defined as
    \begin{equation*}
        H_j = \partial_j h = \Delta U_j \sqrt{\frac{1}{K \sum_{i = 1}^K (\Delta U_i)^2}} = \frac{\Delta U_j}{K \Delta_\mathrm{REO}}.
    \end{equation*}
    Then, under the assumption of delta method, we have
    \begin{equation*}
        \mathrm{Var}(\Delta_\mathrm{REO}) =: \Xi = H^\top \Sigma H.
    \end{equation*}
    Here, $\Sigma_{j,k} := \mathrm{Cov}(\Delta U_j, \Delta U_k)$ is the covariance matrix of relative group utility functions. Let the matrix element of the Jacobian matrix be
    \begin{equation*}
        G_{j, k} := \partial_j g_k := K \frac{\delta_{j, k} \sum_{i = 1}^K U_i - U_k}{(\sum_{i = 1}^K U_i)^2}
    \end{equation*}
    and let the covariance matrix of the group utility function be
    \begin{equation*}
        \mathrm{Cov}(U_j, U_k) =: \Gamma_{j, k} = \delta_{j, k} U_k^2 \left(\frac{1 - q_k}{q_k} \frac{1}{\abs{\mc{D}_{rec}}} + \frac{1 - p_k}{p_k} \frac{1}{\abs{\mc{D}_{rand}}} \right).
    \end{equation*}
    Using delta method again, we have
    \begin{equation*}
        \Sigma = G^\top \Gamma G \quad \text{ and } \quad  \Xi = H^\top G^\top \Gamma G H = \sum_{j = 1}^K \Gamma_{j, j} (G H)_j^2.
    \end{equation*}
    It is worth noting that both $\Sigma$ and $\Xi$ are invariant under the rescaling of the group utility function. Suppose we rescale the group utility function simultaneously as $\wt{U}_k = \alpha U_k$ for any $k = 1, \cdots, K$. Then, the related matrices are rescaled as $\wt{\Gamma} = \alpha^2 \Gamma$, $\wt{G} = \alpha^{-1} G$ and $\wt{H} = H$. Consequently, the rescaling invariance of covariance matrices follows the defining relations
    \begin{equation*}
        \wt{\Sigma} = \wt{G}^\top \wt{\Gamma} \wt{G} = \Sigma \quad \text{ and } \quad \wt{\Xi} = \wt{H}^\top \wt{G}^\top \wt{\Gamma} \wt{G} \wt{H} = \Xi.
    \end{equation*}

\section{Alternative procedures for A/B test}

According to the construction, the denominators of the RTPR utilities in both control and treatment groups are derived from the same dataset $\mathcal{D}_{rand}$. It might introduce unwanted correlations between estimators from two groups. However, notice that the denominator is increasingly close to a scalar constant when the size of the dataset $\abs{\mathcal{D}_{rand}}$ becomes large as suggested by the law of large numbers. Hence, it is reasonable to assume that estimators $\hat{U}_k^C$ and $\hat{U}_k^T$ from control and treatment groups are weakly correlated in the large sample limit in terms of $\abs{\mathcal{D}_{rand}}$. Consequently, in large sample regime, the correlation will be treated as high-order perturbation. Thus, the hypothesis testings in the A/B test fit the framework of two normal-distributed sample tests thanks to the asymptotic normal distribution derived in proofs of \cref{thm:estimate-delta-Uk,thm:estimate-REO}. This leads to two practical algorithms for A/B tests.

\subsection{Partition-based tests}
In practice, the dataset queried from the database is usually of a very large scale. Hence, partitioning the full dataset into a collection of non overlapped subsets can be used to retrieve more statistical information. Given a dataset $(\mathcal{D}_{rec}, \mathcal{D}_{rand})$, a $M$-fold partition gives a collection of subsets $\{(\mathcal{D}_{rec}(j), \mathcal{D}_{rand}(j)) : j \in [M]\}$. Computing REO metric on each subset, we have $\{\wh{\Delta_\mathrm{REO}}(j) : j \in [M]\}$ which are i.i.d. normal distributed due to large-size and non-overlapping assumptions. Considering that the sample size of the control group may differ than that of treatment group, the partition fold number may differ. Let $M_C$ and $M_T$ be the partition fold numbers of dataset from control and treatment groups. The previous procedure gives the mean and standard deviation of subset REO metrics from each group as $\mu_C := \mathrm{mean}(\wh{\Delta_\mathrm{REO}^C}(\cdot))$, $s_C := \mathrm{std}(\wh{\Delta_\mathrm{REO}^C}(\cdot))$, and $\mu_T, s_T$ defined similarly. Hence, Welch's t-test indicates that the pivot statistic follows student t-distribution
\begin{equation*}
    T := \frac{D_\mathrm{REO} - (\mu_T - \mu_C)}{\sqrt{s_T^2 / M_T + s_C^2 / M_C}} \sim t_\nu, \text{ where } \nu \approx \left\lfloor \frac{(s_T^2 / M_T + s_C^2 / M_C)^2}{s_T^4 / (M_T^2(M_T - 1)) + s_C^4 / (M_C^2(M_C - 1))} \right\rfloor.
\end{equation*}
Inverting this pivot distribution gives the confidence interval
\begin{equation}\label{eqn:confidence-interval-partition}
   \mu_T - \mu_C - t_\nu(\delta / 2) \sqrt{s_T^2 / M_T + s_C^2 / M_C} \le D_\mathrm{REO} \le \mu_T - \mu_C + t_\nu(\delta / 2) \sqrt{s_T^2 / M_T + s_C^2 / M_C} 
\end{equation}
where $1 - \delta$ is confidence level.

\begin{algorithm}[htbp]
\caption{Partition-based A/B tests for REO metrics}
\label{alg:practical-AB-test-multiple-sample}
\begin{algorithmic}
\State \textbf{Input: } Confidence level $1 - \delta$, partition fold numbers $M_C, M_T$, dataset $(\mathcal{D}_{rec}^T, \mathcal{D}_{rec}^C, \mathcal{D}_{rand})$.
\State Initialize arrays $\mathbf{v}_C \in \mathbb{R}^{M_C}$ and $\mathbf{v}_T \in \mathbb{R}^{M_T}$.
\State Partition the original dataset into non overlapped subsets $\{(\mathcal{D}_{rec}^C(j), \mathcal{D}_{rand}^C(j)) : j \in [M_C]\}$ and $\{(\mathcal{D}_{rec}^T(j), \mathcal{D}_{rand}^T(j)) : j \in [M_T]\}$.
\For{$j=1,\ldots, M_C$}
   \State Estimate REO metric $\mathbf{v}_C(j) \leftarrow \wh{\Delta_\mathrm{REO}}(j)$ on the $j$-th subset $(\mathcal{D}_{rec}^C(j), \mathcal{D}_{rand}^C(j))$.
\EndFor
\For{$j=1,\ldots, M_T$}
   \State Estimate REO metric $\mathbf{v}_T(j) \leftarrow \wh{\Delta_\mathrm{REO}}(j)$ on the $j$-th subset $(\mathcal{D}_{rec}^T(j), \mathcal{D}_{rand}^T(j))$.
\EndFor
\State Set $\mu_C = \mathrm{mean}(\mathbf{v}_C), s_C = \mathrm{std}(\mathbf{v}_C)$ and $\mu_T = \mathrm{mean}(\mathbf{v}_T), s_T = \mathrm{std}(\mathbf{v}_T)$.
\State \textbf{Output: } Estimation $D_\mathrm{REO} \approx \mu_T - \mu_C$ and confidence interval derived according to \cref{eqn:confidence-interval-partition}.
\end{algorithmic}
\end{algorithm}

\subsection{Bootstrap-based tests}
When the size of dataset is not large enough, partitioning may weaken the representation ability of the statistical information of the dataset. Alternative option for constructing confidence interval is estimating the variance of the difference-in-REO metric by bootstrap. Given a dataset $(\mathcal{D}_{rec}^T, \mathcal{D}_{rec}^C, \mathcal{D}_{rand})$, bootstrap procedure gives a collection of datasets $\{(\mathcal{D}_{rec}^T(j), \mathcal{D}_{rec}^C(j), \mathcal{D}_{rand}(j)) : j \in [B]\}$ of the same size by sampling the original dataset with replacement. Computing difference-in-REO on each bootstrap dataset, a collection of bootstrap metrics $\{\wh{D_\mathrm{REO}}(j) : j \in [B]\}$ is derived. Then, the standard error of difference-in-REO is approximated by the standard deviation of bootstrap metrics $\mathrm{se}(\wh{D_\mathrm{REO}}) \approx \mathrm{std}(\wh{D_\mathrm{REO}}(\cdot))$. Suppose $\wh{D_\mathrm{REO}}$ is the difference-in-REO metric computed on the original dataset. Then, the confidence interval is
\begin{equation}\label{eqn:confidence-interval-bootstrap}
    \wh{D_\mathrm{REO}} - z(\delta / 2) \mathrm{se}(\wh{D_\mathrm{REO}}) \le D_\mathrm{REO} \le \wh{D_\mathrm{REO}} + z(\delta / 2) \mathrm{se}(\wh{D_\mathrm{REO}})
\end{equation}
where $1 - \delta$ is the confidence level and $z(\delta / 2)$ is the quantile of standard normal distribution.

\begin{algorithm}[htbp]
\caption{Bootstrap-based A/B tests for REO metrics}
\label{alg:practical-AB-test-bootstrap}
\begin{algorithmic}
\State \textbf{Input: } Confidence level $1 - \delta$, bootstrap size parameter $B$, dataset $(\mathcal{D}_{rec}^T, \mathcal{D}_{rec}^C, \mathcal{D}_{rand})$.
\State Initialize an array $\mathbf{v} \in \mathbb{R}^B$.
\State Estimate $\wh{D_\mathrm{REO}}$ from the original dataset.
\For{$j=1,\ldots, B$}
   \State Sample bootstrap dataset $(\mathcal{D}_{rec}^T(j), \mathcal{D}_{rec}^C(j), \mathcal{D}_{rand}(j))$ from the original dataset.
   \State Estimate difference-in-REO metric $\mathbf{v}(j) \leftarrow \wh{D_\mathrm{REO}}(j)$ on the $j$-th bootstrap dataset.
\EndFor
\State Set $\mathrm{se}(\wh{D_\mathrm{REO}}) \leftarrow \mathrm{std}(\wh{D_\mathrm{REO}}(\cdot))$.
\State \textbf{Output: } Estimation $D_\mathrm{REO}$ and confidence interval derived according to \cref{eqn:confidence-interval-bootstrap}.
\end{algorithmic}
\end{algorithm}

\section{Discussions on Other Fairness Metrics}

Our work primarily focused on the item-side REO fairness notion in recommendation systems, this is not only because we focus on personalized recommendation scenarios that respect users' diverse preferences, but also because random traffic is not necessary for other popular fairness notions. 

\subsection{RSP Does Not Require User Preference Information}
For item-side fairness in recommendation systems, Ranking-based Statistical Parity (RSP) is another popular notion that does not depend on the users' preferences. Specifically, it is an extension of the Statistical Parity/Demographic Parity fairness notion in binary classification, which aims to use the group-wise Ranking-based Positive Rates (RPR) as utilities $U_k$ and measures the disparities among them, i.e., 
\begin{equation}
    U_k := \frac{\sum_{(u,i) \in \mathcal{D}} \mathbb{I}\{R(u,i) = 1\} \mathbb{I} \{S(i) = s_k\}}{\sum_{(u,i) \in \mathcal{D}} \mathbb{I} \{S(i) = s_k\}},
\end{equation}
\begin{equation}
    \Delta U_k := U_k - \frac{1}{K} \sum_{k'} U_{k'},
\end{equation}
\begin{equation}
    \Delta_\mathrm{RSP} := \frac{\mathrm{std}(U_1, \dots, U_K)}{\mathrm{mean}(U_1, \dots, U_K)}.
\end{equation}
To obtain the denominator $\sum_{(u,i) \in \mathcal{D}} \mathbb{I}_{S(i) = s_k}$, the platform only needs to track the posting behavior of creators and do not need to deliver them to the users to obtain user-interaction labels.

\subsection{Realistic Constrained User-Side RTPR can be Computed without Random Traffic}

For user-side fairness, we note that using precision$@ N_{show}$ as the utility measurement is in fact reasonable, which can be computed solely based on $\mathcal{D}_{rec}$, and $N_{show}$ is the number of delivered items for each request. The main difference between the user-side utility and the creator-side utility is the fact that the $N_{show}$ sets a natural upper bound for the maximum user utility.

More specifically, in actual production, for every user request, $N_{show}$ items are returned and thus we have $\sum_{(u,i) \in \mathcal{D}_{rec}}  \mathbb{I} \{S(u) = s_k\}$ as multiples of $N_{show}$, for a user request from $u$, suppose the items (de-duplicated) returned is denoted as $\mathcal{I}_{ret}$, then the precision@$N_{show}$ is computed as $\frac{\sum_{i \in \mathcal{I}_{ret}} \mathbb{I}\{Y(u,i)=1\}}{N_{show}}$. Aggregating all user requests, we get the user-group utility as 
\begin{equation}
    U_k = \frac{\sum_{(u,i) \in \mathcal{D}_{rec}} \mathbb{I}\{Y(u,i) = 1\} \mathbb{I} \{S(u) = s_k\}}{\sum_{(u,i) \in \mathcal{D}_{rec}}  \mathbb{I} \{S(u) = s_k\}}.
\end{equation}

To explain why precision@$N_{show}$ is a reasonable replacement to Ranking-based TPR, we denote $\mathcal{I}^{seen}_{u}$ as the set of items previously delivered to the user, and define 
\begin{equation*}
    \mathcal{I}^{left}_{u} := \{ i | i \in {\mathcal{I} - \mathcal{I}^{seen}_{u}}, Y(u,i) = 1\},
\end{equation*}
as the set of unseen items to $u$.
Then we note that in general we have $N_{show} < | \mathcal{I}^{left}_{u} |$, i.e., $N_{show} = \min\{N_{show}, | \mathcal{I}^{left}_{u} | \}$ (there are more than $N_{show}$ that are relevant in the candidate pool, otherwise the user has almost no reason to keep using this platform).

\begin{figure}[h!]
    \centering
\includegraphics[width=\linewidth]{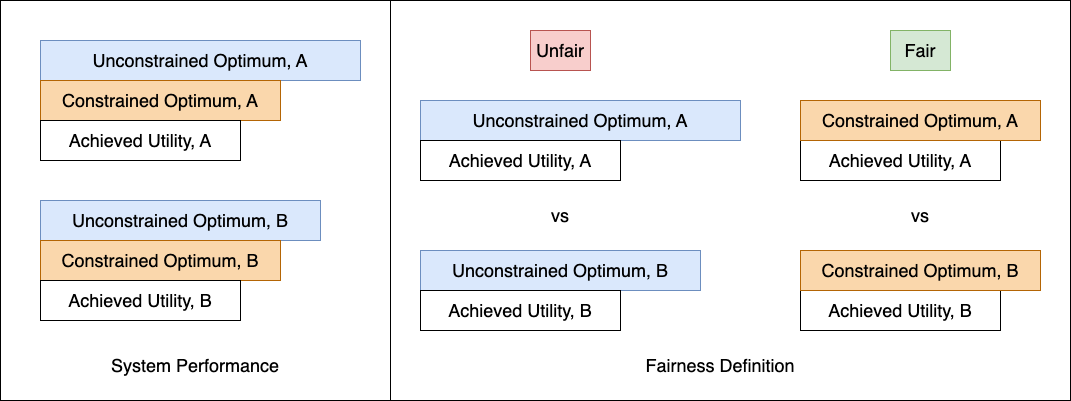}
    \caption{User-side fairness perspectives, when the constrained optimum is used as the upper bound, Precision@$N_{show}$ is a reasonable utility.}
    \label{fig:user_side_fair}
\end{figure}

In the unconstrained situation, the TPR is computed by
\begin{equation*}
    \frac{\sum_{i \in \mathcal{I}_{ret}} \mathbb{I}\{Y(u,i)=1\}}{|\mathcal{I}^{left}_{u}|},
\end{equation*}
But since the system can only present $N_{show}$ items, it is reasonable to have a constrained version of the TPR utility, which is 
\begin{equation*}
    \frac{\sum_{i \in \mathcal{I}_{ret}} \mathbb{I}\{Y(u,i)=1\}}{\min\{N_{show}, | \mathcal{I}^{left}_{u} | \}} = \frac{\sum_{i \in \mathcal{I}_{ret}} \mathbb{I}\{Y(u,i)=1\}}{N_{show}},
\end{equation*}
and thus is equivalent to precision@$N_{show}$, which can be computed solely based on $\mathcal{D}_{rec}$ (see Figure \ref{fig:user_side_fair} for illustrations on the constrained vs unconstrained perspectives on the RTPR utility).

\section{Further Discussions on The Actual Recommendation System} \label{sec:discussion}

In this part, we discuss how we implement the entire REO-related metrics measurement pipeline in real-world recommendation systems 

\subsection{Content De-duplication} \label{appsec:dedup}

An important aspect of real-world recommendation systems is \emph{removing repetitive content} in the recommendations. During online serving, we mark the traffic source of each delivered item, and when the random traffic and the default traffic both recommend an item to the user, we only keep that item once, but dump this user-item pair's data to both $\mathcal{D}_{rec}$ and $\mathcal{D}_{rand}$, e.g., if $i$ is previously recommended in the default traffic and is recommended to $u$ by the random traffic later, we do not recommend it again, but dump the data corresponding to $(u,i)$ to $\mathcal{D}_{rand}$.

\subsection{Volume/Fraction of the Random Traffic}

Another important design parameter is the \emph{volume/fraction} of the random traffic. In actual production, random traffic will ignore the user's preference, and thus too large a random traffic will not expose users sufficiently to their interests and hurt the user experience. 

\paragraph{Tradeoff between recommendation relevance and measurement accuracy.} The platforms can benefit from both a high recommendation relevance and high accuracy of fairness metric measurements, and thus we can write out the platform's utility function as
\begin{equation}
    U_{platform} := (1-\gamma)U_{rel} + \gamma U_{acc}(\epsilon, \delta)
\end{equation}
where $U_{rel} := \beta U^T_{rel} + \beta U^C_{rel} + \alpha U^{rand}_{rel}$ measures the recommendation relevance utility component, and $\beta$, $\alpha$ denote the treatment/control group volume and random traffic volume in this A/B test respectively; $U_{acc}(\epsilon, \delta)$ is determined by $\epsilon$ and $\delta$ in \cref{thm:main_estimate__reo_metrics}, and $\gamma \in (0,1)$ measures the relative importance of the two parts. We note that in general, we have $U^{rand}_{rel} < \min\{U^T_{rel}, U^C_{rel}\}$ and thus increasing $\alpha$ will decrease $U_{rel}$, please see \cref{app:random_traffic_diversity} for further discussion. Meanwhile, it is obvious from our previous analysis in \cref{thm:main_estimate__reo_metrics} that $U_{acc}$ increases in $\alpha$ can make sure the system collects sufficiently large $\mathcal{D}_{rand}$ more quickly for given values of $\epsilon$ and $\delta$, resulting in the tradeoff.

\paragraph{Orthogonal traffic assignment on the same set of users.} A good aspect of orthogonal traffic assignment is that the same random traffic can be used to measure the REO-related metrics for all experiments running on these users so we can avoid further sacrificing recommendation relevance.

\paragraph{Optimal random traffic volume is not universal across different A/B tests.} From \cref{sec:proofs}, we can tell that to achieve certain estimation accuracy $U_{acc}(\epsilon, \delta)$, the required sample size depends on $p_k$ and $q_k$ values, where the $q_k$ values can highly depend on the implemented treatment and control strategies and influence the number of samples needed. While the system can set different random traffic volumes on different sets of users, A/B tests with orthogonal traffic assignment on the same set of users share the same random traffic volume and the system can choose different time window lengths to get the respective data sizes for these tests.

\subsection{Random Traffic and Diversity} \label{app:random_traffic_diversity}

Despite hurting the recommendation relevance in expectation, random traffic is said to help explore the users' new interests and boost recommendation diversity \cite{borgs_bursting_2023, Moller_2018}, so a small volume of random traffic can be beneficial for the platform both in terms of metrics estimation and user experience (our platform actually observes a statistically significant increase in users' daily stay time when increasing the random traffic fraction from 0 to 0.02$\%$). This indicates that the $U_{rel}$ part is not necessarily monotone in the fraction of default strategy, which is reasonable since recommendation models are trying to estimate users' preferences based on historical interactions and are not perfect at characterizing their preferences, especially when certain types of items never appear in the users' history.

\subsection{Stage-wise Fairness Diagnosis}

\begin{figure}[h!]
    \centering
        \includegraphics[width=0.6\linewidth]{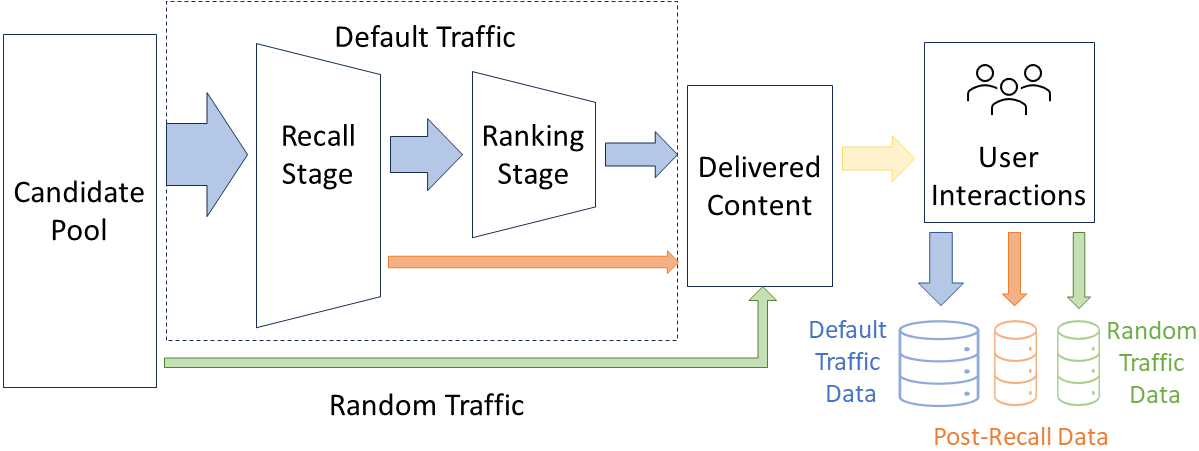}
        \caption{The system's traffic design for stage-wise fairness diagnosis.}
        \label{fig:stagewise}
\end{figure}

We also note that our method can generalize beyond measuring the fairness of the entire recommendation system. For instance, if we want to diagnose if the recall stage is fair in terms of REO, we can force insertion a \emph{post-recall} (or pre-ranking) traffic into the delivered contents and get a set of user-item pairs from it, denoted by $\mathcal{D}_{pr}$ (see \cref{fig:stagewise}). Then if we replace the terms obtained from $\mathcal{D}_{rec}$ with $\mathcal{D}_{pr}$ in previous sections, we can estimate the REO-related metrics of the recall stage. Similarly, if we replace the terms obtained from $\mathcal{D}_{rand}$ with $\mathcal{D}_{pr}$ in previous sections, we can estimate the REO-related metrics of the ranking stage.
\begin{figure*}[h!]
    \centering
    \begin{minipage}[h]{.6\textwidth}
    \includegraphics[width=\linewidth]{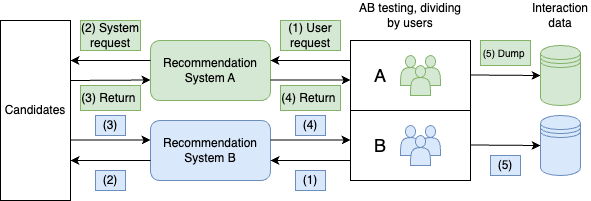}
    \caption{An illustration of A/B testing where users are divided into two groups that receive control (A) and treatment (B) recommendation strategies respectively.}
    \label{fig:ab_testing}
    \end{minipage}
    \hfill
    \begin{minipage}[h]{.36\textwidth}
        \includegraphics[width=\linewidth]{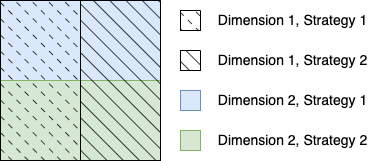}
    \caption{An illustration of orthogonal traffic assignment in A/B testing, it is common for A/B strategies tested on orthogonal dimensions.}
    \label{fig:orthogonal_traffic}
    \end{minipage}
\end{figure*}

\section{Machine Specs}

All of our experiments were conducted on a single machine with an AMD EPYC 9124 CPU and 64 GB memory, without a GPU.

\end{document}